\documentstyle[12pt,aaspp4]{article}

\tighten

\begin{document}
\def\etal{{et al.~}} 
\def\spose#1{\hbox to 0pt{#1\hss}}
\def\simlt{\mathrel{\spose{\lower 3pt\hbox{$\mathchar"218$}}
     \raise 2.0pt\hbox{$\mathchar"13C$}}}
\def\simgt{\mathrel{\spose{\lower 3pt\hbox{$\mathchar"218$}}
     \raise 2.0pt\hbox{$\mathchar"13E$}}}
\def\ltsima{$\; \buildrel < \over \sim \;$}
\def\gtsima{$\; \buildrel > \over \sim \;$}

\title{The Evolution of Early--Type Galaxies in
Distant Clusters\footnote{Based on observations made with the NASA/ESA
Hubble Space Telescope, obtained from the data archive and at the
Space Telescope Science Institute, which is operated by the
Association of Universities for Research in Astronomy, Inc., under
cooperative agreement with the National Science Foundation} }

\author{S.A.\ Stanford\altaffilmark{2}$^,$\altaffilmark{3},} 
\affil{Institute of
Geophysics and Planetary Physics, Lawrence Livermore National
Laboratory, Livermore, CA 94550} 
\authoremail{adam@igpp.llnl.gov}
\altaffiltext{2}{Jet Propulsion Laboratory, California Institute of
Technology} 
\altaffiltext{3}{Visiting Astronomer, Kitt Peak National
Observatory, National Optical Astronomy Observatories, which is
operated by the Association of Universities for Research in Astronomy,
Inc., under cooperative agreement with the National Science
Foundation.}

\author{Peter R.\ Eisenhardt\altaffilmark{3},} \affil{Jet Propulsion
Laboratory, California Institute of Technology, Pasadena, CA 91109}
\authoremail{prme@kromos.jpl.nasa.gov}

\and
\author{Mark Dickinson\altaffilmark{3,4}}
\affil{Department of Physics and Astronomy, The Johns Hopkins University, 
Baltimore, MD 21218}
\altaffiltext{4}{Alan C. Davis Fellow, also with the Space Telescope 
Science Institute}
\authoremail{med@stsci.edu}

Accepted for publication in The Astrophysical Journal.

\begin{abstract} 

We present results from an optical--infrared photometric study of
early--type (E+S0) galaxies in 19 galaxy clusters out to $z = 0.9$.
The galaxy sample is selected on the basis of morphologies determined
from {\it HST} WFPC2 images, and is photometrically defined in the
$K$--band in order to minimize redshift--dependent selection biases.
Using new ground--based photometry in five optical and infrared bands
for each cluster, we examine the evolution of the color--magnitude
relation for early--type cluster galaxies, considering its slope,
intercept, and color scatter around the mean relation.  New
multiwavelength photometry of galaxies in the Coma cluster is used to
provide a baseline sample at $z \approx 0$ with which to compare the
distant clusters.  The optical--IR colors of the early--type cluster
galaxies become bluer with increasing redshift in a manner consistent
with the passive evolution of an old stellar population formed at an
early cosmic epoch.  The degree of color evolution is similar for
clusters at similar redshift and does not depend strongly on the
optical richness or x--ray luminosity of the cluster, suggesting that
the history of early--type galaxies is relatively insensitive to
environment, at least above a certain density threshold.  The slope of
the color--magnitude relationship shows no significant change out to
$z = 0.9$, providing evidence that it arises from a correlation
between galaxy mass and metallicity, not age.  Finally, the intrinsic
scatter in the optical--IR colors of the galaxies is small and nearly
constant with redshift, indicating that the majority of giant,
early--type galaxies in clusters share a common star formation
history, with little perturbation due to uncorrelated episodes of
later star formation.  Taken together, our results are consistent with
models in which most early--type galaxies in rich clusters are old,
formed the majority of their stars at high redshift in a
well--synchronized fashion, and evolved quiescently thereafter.  We
consider several possible effects which may be introduced by the
choice of morphologically recognizable elliptical and S0 galaxies in
dense environments as a subject for study.  In particular, the
inclusion of S0 galaxies, which might be undergoing morphological
transformation in clusters as part of the Butcher--Oemler effect, may
influence the results of our investigation.

\end{abstract}


\lefthead{Stanford, Eisenhardt, and Dickinson}
\righthead{Evolution of Early--Type Cluster Galaxies}

\section{Introduction}

The elliptical galaxy formation scenario proposed by, e.g., Eggen,
Lynden--Bell, \& Sandage (1962, hereafter ELS), Searle, Sargent, \&
Bagnuolo (1973), and Tinsley \& Gunn (1976) postulates a single burst
of star formation at high redshift, followed by passive stellar evolution.
An elliptical galaxy is assumed to form the vast majority of its stellar
mass during the initial starburst.  Several observations suggest that
early--type galaxies in present--day clusters may have formed and evolved
in this fashion.  For example, the color--magnitude ($c-m$) relation seen 
in nearby clusters can be readily explained in terms of a single star 
formation episode.  More massive galaxies would retain supernovae ejecta 
more effectively, resulting in higher metallicities for the succeeding 
generations of stars within the initial burst, and hence in redder colors 
for more luminous galaxies (Larson 1974; Arimoto \& Yoshii 1987; 
Franx \& Illingworth 1990).  Indeed, the $c-m$ relation in nearby clusters
evidently implies a close tie between metallicity and galaxy mass, as
seen in the tightness of the Mg$_2$--$\sigma$ correlation (Bender,
Burstein, \& Faber 1993).  Furthermore, the scatter in the $UVK$
colors of present--epoch cluster E+S0s is observed to be very small
(Bower, Lucey, \& Ellis 1992; Eisenhardt et al.\ 1997).  This has been
used to argue for a high degree of synchronization in their star
formation histories, and for relatively old ages.  Late, episodic bursts 
of star formation, young galaxy ages, or a wide range in galaxy formation 
redshifts would be expected to lead to larger color scatter than is 
observed.  An early formation epoch coupled with passive evolution simply 
predicts the observed homogeneity in the colors of most elliptical galaxies 
in clusters today.  

There is, however, spectroscopic evidence for younger stellar populations 
in some elliptical and S0 galaxies (O'Connell 1980;  Worthey 1996), 
as well as morphological signs of disturbance attributed to past merger 
events (Toomre 1978; Schweizer \& Seitzer 1988).  Moreover, Schweizer 
et al.\ (1990) and Schweizer \& Seitzer (1992) have shown that the 
spectroscopic and photometric indicators of younger starlight correlate 
with the degree of morphological disturbance in E and S0 galaxies, 
implying a connection between the two phenomena.  The galaxy samples 
considered in such studies have consisted primarily of {\it field} 
galaxies, and have not included galaxies in the cores of rich clusters 
such as Coma.  It is therefore unclear whether these results apply to 
all early--type galaxies or are a function of their environment.

The traditional picture of massive galaxy formation in a monolithic
collapse episode at high redshift does not sit comfortably within the
context of modern hierarchical merging scenarios for galaxy formation
and evolution, such as those based on the cold dark matter (CDM)
model.  Such models predict that massive galaxies form late, at $z \le
2$, from the gradual merging of smaller galaxies (e.g., White \& Frenk
1991; Kauffmann et al.\ 1993; Cole et al.\ 1994).  Although this
scenario predicts a wide range of ages for elliptical galaxies, their
small color scatter in the present epoch can be accomodated because
sufficient time elapses since the epoch of extensive merging for the
resulting color variations to damp out (Kauffmann 1996; see also
Schweizer \& Seitzer 1992).  The mass--metallicity correlation implied
by the slope of the color--magnitude relation has been recently
examined in the context of hierarchical merging models by Kauffmann \&
Charlot (1997).  Using a multi--metallicity spectral synthesis code
and semi--analytical galaxy evolution models, they are able to
reproduce this correlation by forming more massive ellipticals from
mergers of more massive progenitor disk galaxies, which themselves are
better able to retain metals during star formation.

The aforementioned observational constraints on E/S0s are well--known
only at $z = 0$, which allows a wide range of possible star formation
histories, including both the ELS and CDM scenarios.  Recent
observational advances both on the ground and in space are beginning
to provide detailed information on the properties of early--type
galaxies at higher redshift, e.g., through study of the Fundamental
Plane and its projections (Dokkum \& Franx 1996; Dickinson 1995 and
1997; Pahre et al.\ 1995), and the Mg--$\sigma$ relation (Ziegler \&
Bender 1997).  By examining the properties of galaxies in clusters at
high redshifts, constraints on the nature of early--type galaxy
evolution might be developed so as to favor one formation scenario
over the other.

In practice, several complications have arisen in investigations of
galaxy evolution in distant clusters.  Photometric studies based on
optical imaging (e.g., Butcher \& Oemler 1978; Dressler \& Gunn 1992;
Rakos \& Schombert 1995; and Lubin 1996) may be hampered by selection
effects due to the redshifting of blue and ultraviolet rest frame
wavelengths into the observed bands.  This is a significant concern
for galaxy selection in the most distant known clusters from optical
and x--ray samples ($z \approx 0.9$), where even the $I$--band
measures blue wavelengths in the cluster rest frame.  Selecting galaxy
samples in the near infrared alleviates this problem.  Even at $z \sim
1$, the $K$--band measures galaxy light emitted in the rest frame
near--IR.  Because the infrared spectral energy distributions of all
but the most vigorously star forming galaxies are very similar (e.g.\
Johnson 1966), a galaxy sample defined in the near--IR should be free
of redshift--dependent biases regarding galaxy type out to $z \approx
1$ and beyond.  Optical--to--infrared photometry, particularly where
the optical band measures light emitted shortward of 4000 \AA\ in the
rest frame, ensures a long wavelength baseline for color measurements,
providing in effect a measurement of the luminosity ratio of main
sequence stars to evolved red giants in a galaxy's stellar population
(e.g. Bruzual \& Charlot 1993).  Arag\'on--Salamanca et al.\ (1993)
used this approach to study color evolution in a study of 10 $z \ge
0.5$ clusters, and found that the modal optical--IR colors of the
galaxies, presumed to be primarily cluster ellipticals, become bluer
with redshift.

Without the ability to distinguish between morphological classes of
galaxies in distant clusters, purely ground--based photometric studies
run the risk of mixing galaxy types.  This results in a particular hazard
for studying the evolution of early--type galaxies, since the
increasing proportion of blue, late--type galaxies in distant clusters
(the ``Butcher--Oemler effect'') may introduce greater contamination 
at higher redshifts.  Imaging data from the Hubble Space Telescope 
({\it HST}) largely solves this problem by providing the capability to 
select galaxies purely by morphology (Dressler et al.\ 1994; Couch et al.\
1994).  In Stanford, Eisenhardt \& Dickinson (1995; henceforth SED95)
we made a first attempt to combine multi--band optical--IR
photometry with {\it HST} imaging of two clusters to evaluate the
evolutionary state of early--type cluster galaxies at $z \approx 0.4$.
In that paper, the rest--frame $V-H$ colors of distant,
morphologically--selected early--type galaxies were found to be similar
($< 0.2$ mag bluer) to those at $z = 0$, and the slope and scatter of
the color--magnitude relation for those galaxies were also found to be
indistinguishable from the present--day values.  Ellis et al.\ (1997)
have recently made a similar test using multi--color optical WFPC2 
data to study E+S0 galaxies in three clusters at $z \approx 0.54$,
with similar results.

In the present paper, we extend the analysis of SED95 to a much larger
sample of galaxy clusters, spanning a wide redshift range ($0 < z <
0.9$), using substantially improved optical--IR photometric data.  For
$\Lambda = 0$ cosmologies with $0.05 < q_0 < 0.5$, our cluster sample
spans 50--62\% of the lookback time to the Big Bang, affording us a
broad view of the evolution of early--type galaxies throughout the
second half of the universe's lifespan.  Our major results are
presented here; the photometric data set itself will be presented in a
supplemental paper (Stanford et al.\ 1997).  Except
where noted, the assumed cosmology is $H_0 = 65$ km s$^{-1}$
Mpc$^{-1}$, $q_0 = 0.05$, and $\Lambda = 0$, which results in a
present--day age for the Universe of 13.5~Gyr.

In this paper, we use the term ``early--type'' galaxy to refer to
those galaxies classified morphologically as having Hubble classes E,
E/S0, or S0.  We have made no attempt here to further subdivide the
early--type galaxy population in high redshift clusters, and in
particular no effort has been made to separate elliptical from S0
galaxies.  Making such morphological distinctions is sometimes
difficult even for nearby galaxies, and is dependent on orientation: a
face--on S0, for example, can be hard to distinguish from a
``true'' elliptical galaxy.  It has yet to be established how well
this subclassification can be achieved for WFPC2 images of high
redshift galaxies (cf. Smail et al.\ 1997 for a discussion), and for
the purposes of this paper we have preferred to avoid the resulting
uncertainties which might arise from misclassification.  However,
grouping elliptical and S0 galaxies together for an evolutionary study
such as this one may have undesirable consequences for the interpretation of the
results.  We will return to this point in \S4.  Throughout the text,
except where we wish to make the distinction between elliptical and S0
galaxies explicit, we will generally use the terms ``early--type'' and
E+S0 interchangeably to refer to the overall population of elliptical
and S0 galaxies in a cluster.

In addition, we note here that the subject of our investigation is
{\it giant} galaxies, not dwarf ellipticals or ``spheroidals.''  As
described in \S2, our photometric data and the analysis thereof are
limited to galaxies not more than $\sim 2$ magnitudes fainter than
present--day (unevolved) $L^{\ast}$ at the redshift of each cluster
examined.  For objects with the typical colors of E+S0s, we are
therefore discussing galaxies with $M_B \simlt -18.5$ (for our adopted
cosmology).

Finally, the results reported here concern only cluster galaxies, 
and do not necessarily bear on the evolution of field ellipticals.
Moreover, because we use WFPC2 images which cover only the core regions
of the distant clusters, we cannot address the evolution of galaxies
located at larger cluster--centric radii.  Some studies of
nearby clusters indicate that early--type galaxies located outside of
the core regions may have experienced recent star formation (Caldwell
et al.\ 1993; Caldwell \& Rose 1997).  Almost all of these objects seem
to be S0s.

\section{Observations, Reductions, and Photometry}

The overall sample of clusters for which we have collected photometric
data is large and heterogeneous, consisting of 46 clusters (as of the
time of this writing) at $0 < z < 0.9$, drawn from a variety of
optical, x--ray, and radio selected samples.  In the present paper, we
restrict our attention to a subsample of 19 clusters for which WFPC2
images were available to us.  The redshift distributions of both the whole 
sample and the {\it HST} subsample are shown in Figure 1.

Optical and near--IR images of the clusters were obtained using CCD
and HgCdTe array cameras on NOAO telescopes at Kitt Peak and Cerro
Tololo in 1993--1996.  This imaging provides multiwavelength
photometric data of uniform quality through a standard set of filters,
using well--understood photometric systems.  These features are
particularly advantageous when attempting to systematically
investigate the evolution of large samples of faint galaxies over a
broad redshift range.  The sample of clusters studied here, along with
the bandpasses used, is summarized in Table~1.  In the near infrared,
atmospheric transmission windows require us to use fixed bandpasses,
so we have imaged through standard $J$, $H$ and $K_s$ filters.
Therefore the infrared bands probe different rest--frame wavelengths
for clusters at different redshifts.

The optical imaging was generally obtained in two bands which we have
adjusted according to the cluster redshift in order to ensure that
they span the $\lambda_0 \sim 4000$\AA\ region in the cluster rest
frame.  The clusters were divided into three redshift ranges for this
purpose: $0.3 < z < 0.45$ ($g$ and $R$ bands); $0.45 < z < 0.7$ ($V$
and $I$ bands); and $0.7 < z < 0.9$ ($R$ and $I$ bands).  Our optical
data set contains Gunn $i$-band images of two clusters (MS 1054.5-0321
and GHO 1322+3027), obtained at the Palomar 5.08m telescope in
February 1995.  Photometry for data taken in the Gunn--Thuan filters
were transformed to the Landolt system using observations of
spectrophotometric standards.  Also, for 3C 220.1 we have used our
WFPC2 F555W and F814W images to obtain $VI$ photometry (after blurring
the WFPC2 images to the seeing of our $K$ image), using the
calibrations given by Holtzman et al.\ (1995).  For two clusters,
Abells 370 and 851, we do not have the desired $g$ band data.  (Also it
should be noted that the $JHK$ data on A370 and A851 were obtained
with PtSi detectors, as reported in SED95.) In this paper, we will
generally refer to the two optical passbands as {\it blue} and {\it
red}.  The former measures rest--frame emission at wavelengths similar
to or somewhat bluer than the rest--frame $U$--band, while the latter
corresponds to rest--frame wavelengths roughly from $B$ to $V$.

Exposure times in all bandpasses were chosen to provide galaxy
photometry with $S/N > 5$ for galaxies with the spectral energy
distributions of present--day ellipticals, down to $\sim$2 magnitudes
fainter than $L^\ast$ (unevolved) at the cluster redshift.  Table 1
lists the no-evolution $K^* + 2$ magnitude ($K_{lim}$) for each cluster.
This permits us to study galaxy properties over a similar range of
luminosities for all clusters in our sample, regardless of their
redshift.  The exact correspondence of the apparent magnitude range
observed for a particular cluster to the luminosities of present--day
galaxies depends both on $q_0$ and on the degree of luminosity
evolution in the galaxy population.  Our images typically cover a
field size of $\sim$1 Mpc at the cluster redshift, which is generally
larger than that covered by the WFPC2 data used to select the
early--type galaxy subsamples (see below).  The optical data were
calibrated onto the Landolt system wherein Vega has $m_V = +0.03$, and
the IR images onto the CIT system wherein Vega has $m = 0$.  The
typical rms of the transformations is 0.02 in the optical and 0.03 in
the near--IR.  The effective angular resolution of the images is
generally limited by the large pixel scale of the infrared arrays, and
is $\sim$1.7 arcsec for the $z < 0.6$ clusters and $\sim$1.2 arcsec
for the more distant objects.

Images in all bandpasses were co--aligned and convolved to matching
point spread functions (PSFs) in order to ensure uniform photometry at
all wavelengths through fixed apertures.  Object detection was carried
out on the $K$ images using a modified version (Adelberger, personal
communication) of FOCAS (Valdes 1982), which was also used to obtain
photometry in each band through circular apertures with a diameter
equal to the twice the PSF FWHM.  The ``$N_{samp}$(total)'' column in
Table 1 gives the total number of galaxies brighter than $K_{lim}$
detected in the area of a WFPC2 field within our $K$ images.  In
calculating $K_{lim}$ we have corrected for the tendency of FOCAS to
measure ``total'' magnitudes which are somewhat fainter than the
``true'' total magnitude.  So the $K_{lim}$ listed in Table 1 are
directly comparable to the photometry displayed in Figure 2 (see
below).  The observing methods, data reduction techniques, and
photometric methods for our ground--based data set are described in
detail in SED95.  All photometry has been corrected for reddening
using the interstellar extinction curve given in Mathis (1990), with
values for $E(B-V)$ taken from Burstein \& Heiles (1982).

Morphological selection of E+S0 galaxies was done on the basis of {\it
HST} WFPC2 images, except for Abell~370 where pre--refurbishment WF/PC
data were used with galaxy classifications kindly provided to us by
A.\ Oemler.  The WFPC2 data were drawn from the {\it HST} archive,
excepting MS~1054.5-032, for which the images were kindly provided by
M.\ Donahue, and 3C~220.1, which is from our own imaging program
(Dickinson \& Broadhurst, in preparation).  For each cluster, separate
exposures reduced by the STScI pipeline procedure were brought into
registration as necessary, and combined with iterative rejection to
remove cosmic rays and pixel defects.  The exposure times ranged from
3 orbits at the lowest redshifts up to 16 orbits on the most distant
cluster.  Morphological classification was performed independently by
two of the authors on the summed WFPC2 images, which were usually in
the F702W or F814W filters (using the reddest band where more than one
was available).  Objects were classified in the following groups:
E/S0, Sa/b, Sc/d, Irr, disturbed/interacting galaxies, or stars.  We
do not attempt here to distinguish between E and S0 galaxies, so the
analysis presented in this paper considers these galaxy types together
as a single class.  Comparison between the classifications of the two
authors showed agreement (within the same group) for $\sim$75\% of the
objects.  For the remaining $\sim$25\%, nearly all of which had types
differing by one group, a consensus was reached.  Because the typing
was done primarily for fairly bright galaxies (e.g., to $I \sim 21.7$
for the $z \sim 0.5$ clusters), and because it is used here only to
distinguish early--type galaxies from all other types,
misclassification is unlikely to be an important problem.  The number
of E+S0 galaxies galaxies brighter than $K_{lim}$ identified in each
cluster is given in the ``N$_{samp}$(E/S0)'' column in Table~1.  In 3C
34, the available WFPC2 data was too shallow to allow morphological
typing to reach the desired $K$ magnitude limit; the typing was
performed to only $K^\ast + 1$ in this case.

\section{Evolution of the Color Magnitude Sequence}

We have chosen to characterize our results by focusing on the {\it
average} properties of the early--type galaxies in each of the
clusters, as represented by the intercept, slope, and scatter of the
galaxy color--magnitude relation in four observed--frame colors:
$blue-K$, $red-K$, $J-K$, and $H-K$. Figure~2 shows these $c-m$
diagrams for 17 of the clusters in our $HST$ subsample (those for A370
and A851 having been previously published in SED95).  The early--type
galaxies selected from the WFPC2 images are indicated by the solid
circles.  A tight color--magnitude sequence of the {\it HST}--selected
E+S0 galaxies is readily visible in Figure 2, demonstrating the broad
similarity of their spectrophotometric properties to those of
early--type galaxies in nearby clusters.  Evolutionary effects are
measured by referencing these color--magnitude relations to those
observed in the low redshift ($z = 0.023$) Coma cluster and
transforming the latter to the high redshift frame.  The transformed
Coma color--magnitude sequences in each color combination are
represented by the dotted lines in each panel.  These transformations
are described in \S 3.1.  The intercept, slope, and scatter in the
high redshift E+S0 color-magnitude relations are presented in \S 3.2,
3.3 and 3.4 respectively.

\subsection{Transformation of the Coma Color-Magnitude Relations}

To the extent practical, we compare the photometry of distant E+S0
cluster galaxies to that of their low redshift counterparts observed
at similar rest--frame wavelengths.  By making this comparison as
directly as possible, we minimize the uncertainties in
$k$--corrections and the dependence on spectral models.  Our method
follows that developed by e.g.\ Lilly (1987) and Arag\'on--Salamanca
et al.\ (1993), and is described in detail in SED95.  We have obtained
new images covering the central $\sim$0.8 Mpc of the Coma cluster in
the $UBVRIzJHK_s$ bands down to $\sim$5 mag below $L^\ast$ (Eisenhardt
et al.\ 1997).  The areal coverage of our Coma sample is similar to
that of the distant clusters.  This data set allows us to estimate the
expected ``no--evolution'' colors in the observed bands for any
distant cluster by interpolation among the observed bands on Coma.  In
other words, for each of our observed distant clusters, we determine
the colors that E+S0 galaxies in Coma would appear to have if they
were moved out to the relevant redshift and observed with the same
filters used for the distant cluster.

To calculate the small $k$--corrections necessary for the
interpolation, the spectrum of an 11 Gyr old model galaxy was
constructed from the Bruzual \& Charlot (1996; henceforth BC96)
population synthesis code.  The BC96 models are generated from
libraries of model atmospheres for stars with various metallicities.
Our template presumes a 1 Gyr starburst followed by passive aging, a
Salpeter IMF, and solar metallicity.  This model provides an adequate
fit to the average $UBVRIJHK$ colors of the Coma early--type galaxies
at $\sim$$L^\ast$, and to a composite optical--IR spectrum (M. Rieke,
personal communication) made from observations of M32 and the M31
bulge.  In the near infrared, the agreement between the BC96 template
and the Rieke spectrum is substantially improved compared to spectra
generated using earlier versions of the Bruzual \& Charlot code,
particularly at $\lambda \approx 1 \mu$m.  Because this template is
only used for interpolation between adjacent bandpasses, the resulting
uncertainties in the transformed Coma photometry should be small.

Finally, for each distant cluster, a linear fit was made to the
transformed Coma photometry in color--magnitude space for each of the
four observed--frame colors: $blue-K$, $red-K$, $J-K$, and $H-K$. 
These no-evolution color-magnitude relations are shown by the 
dotted lines in Figure~2.

\subsection{Color Evolution}

Color evolution in each of the 19 distant clusters was assessed by
measuring the {\it differences} between the transformed Coma $c-m$
relation and the colors of the early--type galaxies with $K < K_{lim}$
selected from the WFPC2 images.  The average color offset was
determined in two ways.  The biweight location estimator (Beers,
Flynn, \& Gebhardt 1990) was calculated using software kindly provided
by T.\ Bird.  A more traditional method was also tested in which an
average (weighted by $K$--band flux) was calculated with one iteration
of 3$\sigma$ clipping.  In this procedure, the average and its
dispersion are calculated, color differences greater than 3$\sigma$
from the average are removed, and then the average is calculated
again.  The main reason for employing the clipping is to reject
galaxies which lie at redshifts different from the cluster.  Because
of the strong $k$--correction to the observed frame colors of distant
ellipticals, foreground and background galaxies should appear to have
substantially different colors from those of the main cluster locus.
The biweight estimator is relatively stable against outliers.  The two
methods gave similar results for the average color differences.

Due to the large number of early--type galaxies in most of the clusters, 
the random photometric errors in the average color differences computed 
here are small compared with the potential systematic errors.  Combining 
in quadrature the uncertainties in the Coma photometric zeropoints, the
distant cluster zeropoints, the extinction correction, PSF matching
errors, color gradient corrections, and $k$--corrections, we estimate
that the total systematic error is $\sim$0.06 mag in a color difference
(see SED95 for a more thorough discussion).

The average color differences for the 19 clusters in the four observed
colors are plotted against redshift in Figure 3, where the error bars
are the uncertainties ({\it not} the scatter), including the estimated
systematic errors.  The horizontal dotted lines in Figure 3 at zero
color difference represent no color evolution relative to Coma.  A
trend toward bluer rest--frame colors at higher redshifts is readily
apparent in the $blue - K$ and $red - K$ data, and even to a small
extent in the $J - K$ data.  The degree of color evolution is greater
for the bluer rest--frame bands, as would be expected if more light
from younger stars contributes to those bands at larger look back
times.

The reliability of the color trends seen in Figure 3 depends to a
large extent on the question of cluster membership.  Considering {\it
all} of the galaxies within the $K < K_{lim}$ samples, clearly many of the
objects in the $c-m$ diagrams of, e.g.,\ GHO 1603+4313 in Fig.\ 2(q)
are not cluster members.  While spectroscopic redshifts have been
measured for some galaxies in this and other clusters, they are
generally relatively few in number, and in many cases they remain
unpublished.  Field contamination of the early--type galaxy samples
selected from within the WFPC2 fields, however, should be
comparatively low.  At both low and high redshifts, elliptical and S0
galaxies dominate the core regions of rich clusters (Dressler 1980;
Oemler et al.\ 1997).  Because the small WFPC2 fields primarily sample
the cluster cores, the contrast of cluster E+S0s over the field
population should be substantially enhanced relative to the overall
galaxy sample.  Therefore we have elected {\it not} to correct for
field galaxy contamination in a statistical sense.

The level of field galaxy pollution in a WFPC2 frame can be estimated
by making use of published number counts and morphological fractions
for the field galaxy population.  In Table~1, N$_{pred}$(field) gives
the predicted number of field galaxies within the WFPC2 area down to
$K_{lim}$, using the $K$ band number counts reported in Djorgovski et
al.\ (1995).  N$_{pred}$(E/S0) is the expected number of field E+S0s,
based on N$_{pred}$(field) and the fraction of early--type galaxies as
a function of ``equivalent'' $\sim$$I$ magnitude given by Driver et
al.\ (1995).  The $I$ magnitude used is that corresponding to the
observed $c-m$ relation at $K_{lim}$.  (While it would be preferable
to use morphological fractions as a function of $K$ magnitude, such
data are not yet available.)  Clearly, the predicted field E/S0
contamination is uncertain, given the small volumes under
consideration in each cluster and our inadequate knowledge of field
galaxy morphological fractions at faint magnitudes.  Taken at face
value, the estimates suggest the early--type galaxy samples in most of
our clusters are dominated by members, and hence the modal properties
of the entire E+S0 samples are representative of the properties of the
true cluster galaxies.  Figure 2 shows that the color trends we
observe are characteristic of our sample and cannot be attributed to
contaminating foreground galaxies.  The relative fraction of
early--type field galaxies may well be greater in the higher redshift
clusters, but this is difficult to assess given the widely varying
numbers seen in the clusters at $z > 0.7$.  Nevertheless, nearly all
of the early--type galaxies in the three highest redshift clusters are
found to be bluer in $R-K$ than the predicted no--evolution line,
indicating that the cluster members alone have bluer rest--frame
colors than do cluster E+S0 galaxies today.

As a further check on contamination, statistical field corrections were
applied to the complete $c-m$ diagram data (without reference to galaxy
morphology) in a manner similar to that described in SED95 but with a
much larger and more homogeneous photometric data set on faint field
galaxies (Elston, Eisenhardt, \& Stanford 1997).  Because of the strong
effects of the $k$--corrections for early--type galaxies, the color
distribution of field ellipticals and S0s (which should span a wide
range of redshifts) is expected to be broader than that of the cluster
E+S0s, and in practice, several of the morphologically--selected
early--type galaxies with colors far from the no--evolution locus are
removed by the statistical procedure.  Average color differences of
these field--corrected early--type samples for each cluster were
calculated and found to be the same within the errors as those
presented in Figure 3.  It should be emphasized that the field
corrections were applied only as a test of our results,
which were calculated from the samples {\it without} field correction.

Other important issues bearing on the observed color evolution
include zeropoint errors and the transformation of the Coma colors.
Even if the zeropoint errors are as much as 0.1 mag (a factor of
$\sim$4 higher than we believe to be the case), which would render the
color differences at low redshift questionable, the higher redshift
optical--$K$ color differences would remain significant.  As for the
Coma--based no--evolution reference colors we calculate, the
transformations should be accurate because they are based on {\it
interpolations} within our own Coma photometry, which itself is in
good agreement with published Coma data such as those of Bower, Lucey,
\& Ellis (1992), and Frogel et al.\ (1978).

\subsection{Color--Magnitude Slopes}

In all 19 of the clusters in our sample, a significant slope to the
optical--IR color--magnitude relationship is evident among the
early--type galaxies.  As for present--day E+S0 galaxies (e.g.,
Sandage \& Visvanathan 1978, De Propris et al. 1997), the slope of the
relation is progressively steeper for colors measured at shorter
wavelengths.  We have fit the slopes of the color magnitude relation
for each of our clusters in various bandpass combinations.  For each
cluster, the $c-m$ fit was made to E+S0 galaxies brighter than
$K_{lim}$, i.e.  over the same range of absolute magnitudes relative
to our assumed no--evolution value of $K^\ast$.  Here it is important
to restate that the optical--IR colors we are observing are {\it not}
fixed in the cluster rest frame: the optical bands shift approximately
with the cluster redshift, while the infrared bands remain fixed.
Therefore the wavelength baseline probed by, e.g., the $blue-K$ color
becomes shorter toward higher redshifts because of the fixed
$K$--band.  This, however, should have a relatively small effect on
interpretation of the slopes measured here, since the intrinsic $c-m$
slope for IR--IR colors is quite small.  As a result, there is little
difference in the slopes of the (e.g.) rest--frame $U-J$ and $U-K$
color--magnitude relations: the bulk of the slope in the
color--magnitude relation comes from the blue rest frame stellar
light.

In Figure 4 we plot the {\it difference} between the measured $blue-K$
versus $K$ slopes and the slope of the Coma cluster E+S0 galaxies at
the same rest--frame wavelengths, evaluated from our transformed Coma
photometry in the same fashion as described in \S3.1 above.  A
negative point in Fig.\ 4 indicates that the cluster has a steeper
slope than does Coma.  The $c-m$ slopes are consistent with those in
Coma, suggesting that little or no {\it differential} color evolution
(as a function of luminosity) has occurred in the early--type galaxy
populations.  Evidently, to the accuracy that we can measure, the
color evolution of giant, early--type cluster galaxies has been
independent of luminosity, down to about 2 magnitudes below $L^\ast$.
We further discuss the implications below.

\subsection{Scatter in the E+S0 Colors} 

As described in the introduction, the small scatter observed for
early--type galaxy colors around the mean locus of the
color--magnitude relation has been used to argue that elliptical
galaxies in the Coma and Virgo clusters have had well--synchronized
star formation histories (Bower, Lucey, \& Ellis 1992).  At higher redshifts,
this test has been made by SED95 using rest--frame $V - H$ colors in
two clusters at $z \approx 0.4$, and by Ellis \etal (1997) using
rest--frame $U-V$ colors in 3 clusters at $z \approx 0.54$.  Here, we
may apply the test to our sample of 19 clusters using a variety of
color combinations over a broad interval of lookback time, from $0 < z
< 0.9$.  As described below, we restrict the sample to galaxies
somewhat brighter than $K_{lim}$ to measure the scatter more
effectively.

Color scatters for the cluster E+S0 samples were calculated using the
mean differences from the transformed Coma relation calculated in \S
3.2 and plotted in Figure~3.  The slope of the color--magnitude
relation was assumed to be fixed at the Coma value for the appropriate
bands, and the scatter about this mean relation was then computed.  As
is evident from Figure~4, the assumption of the Coma $c-m$ slope is a
reasonable one. Any error introduced by the assumption of an incorrect
slope would tend to increase the scatter we measure relative to its
true value.  For each cluster, the observed color scatter and its
uncertainty were calculated using the biweight scale estimator.  The
observed scatter includes contributions from both an {\it intrinsic}
component of color scatter (real color variations among galaxies at a
fixed luminosity) and from photometric uncertainty, which we will call
the {\it measurement} scatter.  In order to determine the level of the
intrinsic component, we first estimate the measurement scatter and
then subtract it in quadrature from the observed value.  Measurement
errors for individual galaxies are larger for fainter galaxies, and
this must be taken into account when computing the overall measurement
scatter expected for the cluster.  Because less luminous galaxies are
more numerous than bright ones, the overall measurement scatter tends
to be dominated by the photometric errors associated with the fainter
galaxies.

The photometric measurement uncertainty expected for each individual
galaxy was determined using a calculated estimate of the
signal--to--noise based on the measured sky background noise and the
galaxy photometry, in conjunction with simulations in which artificial
galaxy images were added to the data at the appropriate magnitudes and
colors and then measured with FOCAS.  This procedure was described in
SED95, and will be elaborated in Stanford et al.\ 1997.  Next, a
series of Monte Carlo simulations was carried out to estimate the
total measurement error over a given magnitude range used in the
estimate of the color--magnitude scatter.  For each cluster, a list of
galaxies was generated matching the actual $K$ magnitudes of the
observed E+S0s for that cluster.  Colors were then assigned to each
galaxy assuming {\it zero} intrinsic color variation plus random
Gaussian noise appropriate to the magnitude of the galaxy.  The
artificial catalog was then run through the biweight estimator to
compute its scatter.  This process was repeated 1000 times for each
cluster and the results were averaged to provide a robust
determination of the expected measurement scatter.

Because, as noted above, photometric errors are larger for fainter
galaxies, there is an optimal magnitude threshold down to which one
should compute the $c-m$ scatter.  Using galaxies which are too faint
inflates the measurement scatter.  At the opposite extreme, using only
a few bright galaxies leaves too few objects with which to determine
the rms color variations, so that the biweight estimator provides only
an uncertain determination of the observed scatter.  The Monte Carlo
simulations described above were repeated with various faint--end
magnitude cutoffs to determine magnitude thresholds at which the
cluster galaxy samples should be limited.  For the low redshift
clusters ($z < 0.5$), where the data generally reaches somewhat deeper
in the rest--frame, this limit was set to $K \leq K^\ast + 1$.  For
clusters at $0.5 < z < 0.7$ it was set to $K \leq K^\ast + 0.5$, and
for clusters at $z > 0.7$ the limit was set to $K^\ast$.  The assumed
value of $K^\ast$ at each redshift assumes $q_0 = 0.05$ and no
luminosity evolution of the galaxy population.  If allowance is made
for either passive luminosity evolution or a larger value of
$q_0$, the scatter calculations cover approximately the same range of
luminosities at all redshifts.  The varying sample depth would affect
the estimated scatter only if the intrinsic scatter changes as a
function of galaxy magnitude (see below).

The observed scatter and the estimates of the measurement scatter in
the four colors spanning the observed wavelength range of our data are
plotted for each cluster in Figure 5a.  Using the subsample of
early--type galaxies truncated to the appropriate $K$ magnitude limit,
the measurement scatter was then subtracted from the observed scatter
in quadrature.  The resulting estimates of the intrinsic rms color
scatter, together with their uncertainties, are plotted in Figure 5b.
To facilitate comparison with similar results from optically--based
studies (e.g.\ Ellis et al.\ 1997), the intrinsic scatter in the
purely optical $blue-red$ color (very roughly $U - V$ in the
rest--frame) is shown in Figure 6.  For some clusters and bandpass
combinations, only upper limits to the intrinsic scatter could be
determined reliably.  Note that implicit in our method of calculation
is the assumption that the intrinsic color scatter in the cluster
galaxies is independent of luminosity within a particular cluster.  We
do not test this here, but Bower, Lucey, \& Ellis (1992) indicate that
this is indeed the case for Coma and Virgo galaxies, while Ellis et
al.\ (1997) suggest that it is true for rest--frame $U-V$ colors of
early-type galaxies in $z \approx 0.5$ clusters as well.

The intrinsic color scatter shown in Figures 5 and 6 is remarkably
constant with redshift in all colors.  No significant trends toward
increasing scatter at higher redshifts are observed.  The dashed line
in each panel shows the intrinsic scatter in a similar rest frame
color for the Coma E+S0 sample -- our measurements of this scatter
from the new Coma data are consistent with those reported by Bower,
Lucey, \& Ellis (1992).  These color dispersions are independent of
many of the uncertainties affecting the colors themselves, notably the
photometric zeropoints, reddening corrections, seeing corrections, and
color gradient corrections.  To the extent that the lack of a field
correction has caused non--cluster members to be included in the
sample, the true scatter may be even lower than the calculated value.
However, the constancy of the intrinsic scatter with redshift argues
against field galaxy contamination being a significant effect.

\section{Discussion}

The overall picture that emerges from the data presented above is one
of generally quiescent behavior for most early--type galaxies in
clusters over at least half the age of the universe.  The
spectrophotometric evolution of the {\it majority} of relatively
bright (and presumably massive) early--type galaxies in clusters over
this time span appears to be smooth and steady.  The parameters of the
color--magnitude relation are very stable out to $z \approx 0.9$:
while the intercept (i.e. the average E+S0 galaxy color at a
particular luminosity) evolves gradually, both the slope and the
scatter of galaxy colors around the mean relation change hardly at
all.

By examining the evolution in the intercept, slope, and scatter of the
galaxy color--magnitude relation, we are describing the {\it average} 
properties of early--type galaxies in rich clusters as a function of 
redshift.  Therefore the evidence that the evolution in these photometric
properties has been ``smooth and steady'' does not exclude the possibility 
that {\it some} early--type cluster galaxies have followed different
evolutionary trajectories, forming later or undergoing later episodes 
of star formation.  Charlot \& Silk (1994) and other authors have noted
that a small amount of late star formation superimposed on an otherwise
old elliptical galaxy leaves only a small imprint on its colors after
approximately 1~Gyr.  Therefore, individual galaxies in the clusters 
studied here may have experienced small, late episodes of star 
formation without standing out too far from the $c-m$ locus.  Indeed, 
if a few extremely blue (or red) ellipticals were present in the cluster, 
these would not strongly perturb the biweight scale estimator used to
determine the color scatter.  

It is also conceivable that the way in which we have constructed our
sample has created a sort of ``tunnel vision'' which systematically
excludes the actively evolving predecessors of early--type cluster
galaxies in today's universe.  We conclude the discussion by considering
the possible role of selection effects on our results.

\subsection{Color Evolution}

The trends of the optical--$K$ colors with redshift presented in \S
3.2 and Figure 3 clearly demonstrate that early--type galaxies in
clusters have evolved since $z \sim 1$.  They are in broad agreement
with similar results from Dressler \& Gunn (1990), Arag\'on--Salamanca
et al.\ (1993), Rakos \& Schombert (1995), and Lubin (1996).  Such a
bluing trend is expected from passive evolution of a stellar
population.  As one approaches younger ages at higher redshifts, the
color of the main sequence turn--off for the bulk of the stellar
population becomes bluer.  The IR--IR colors show little change with
redshift, as expected since the near--infrared light from early--type
galaxies is dominated by giant stars with little contribution from the
main sequence turn--off population.  At the higher redshifts, however,
even the $J-K$ colors have become bluer, in part because the observed
$J$--band samples the rest--frame $R$--band by $z \approx 0.9$.

Some comparison of the observed evolution with models is instructive,
if not definitive.  If early--type cluster galaxies were genuinely a
uniform population of well--synchronized galaxies evolving in a purely
passive fashion, then it would in principle be possible (with
sufficiently accurate data) to compare their spectrophotometric
properties to predictions from reliable population synthesis models,
so as to jointly constrain the evolutionary histories (e.g.\ formation
redshift and star formation timescale) and perhaps the cosmological
parameters (which set the relation between redshift and time).  In
practice, at this stage it is unwise to push such comparisons too far:
the data have their limitations and so do existing models
(cf. Charlot, Worthey \& Bressan 1996).  Moreover, the number of
adjustable parameters is large.  Here we choose to use the models
simply to demonstrate the expected behavior from broad classes of star
formation histories, rather than attempt to uniquely specify
acceptable models from our data.

In Figure 7, we repeat the E+S0 color measurements from Figure 3, 
extending the redshift axis to $z=0$ to include Coma, which lies at 
zero color difference by definition.  In each panel of Figure 7, we 
show several tracks representing predictions for color evolution from 
the spectral synthesis models of Bruzual \& Charlot (1996).  As with 
the data, the model curves represent color {\it differences} relative 
to present--day Coma galaxies, and were calculated in the same way as 
were the color differences of the data.  (The occasional sharp inflections
in the model color tracks, e.g.\ for $blue - K$ at $z=0.7$, reflect the
changes with $z$ in the filter bandpasses used for the optical data.)
In the models, the star formation occurs in a one Gyr burst governed
by a Salpeter IMF, and evolves passively thereafter.  For one of the
cosmologies we have computed models for two formation redshifts, 
$z_f = 2$ and 5, where ``formation'' is defined as the redshift at which
star formation begins.  For our adopted cosmology of $q_0 = 0.05$ and
$H_0 = 65$~km~s$^{-1}$ Mpc$^{-1}$, star formation in these two models
ceased at $z = 1.5$ and 3.1, and the resulting present--day ($z = 0$)
galaxy ages would be 9.7 and 11.9 Gyr, respectively.  We may broadly
consider the two models with differing $z_f$ to represent ``late'' 
and ``early'' formation epochs for the bulk of the stars in 
early--type cluster galaxies.

For the optical--IR colors, particularly the $blue - K$ data, the
``late'' formation models are substantially bluer than the observed
galaxy colors at $z > 0.6$.  For these particular population synthesis
models, this corresponds to a limit on the mean ages of early--type
galaxies in high redshift clusters: model ages younger than
$\sim$4~Gyrs at $z=0.6$ or $\sim$2.7~Gyrs at $z=0.9$ produce colors
too blue to match the actual cluster galaxies.  Also plotted in Fig.\
7 are two other BC96 models where the formation redshift is fixed at
$z_f = 5$ and the cosmology (and hence galaxy age at a given redshift)
is allowed to vary.  There is no strong preference for one of the
cosmologies with the higher $z_f$ shown in Fig.\ 7 if allowance is
made for small zeropoint shifts in the model tracks.

One curious discrepancy for all clusters occurs in the $H-K$ colors.
At all redshifts, the clusters are systematically bluer than the
no--evolution Coma expectation by approximately 0.1~mag.  The same
effect was noted for two clusters at $z \approx 0.4$ in SED95.  The
population synthesis models predict very little evolution in this
color out to $z \approx 0.6$, so the effect is difficult to
understand.  In the rest frame, the $H$--band passes from
$\sim$1.25$\mu$m to $\sim$1$\mu$m in this redshift range -- a part of
the spectral energy distribution of present--epoch galaxies known
poorly at best.  Therefore the observed $H-K$ color at $z \le 0.6$ is
sliding through one of the most uncertain spectral regions of spectral
synthesis models.  However, it is important to recall that the
$\Delta(H-K)$ measurements presented for the data are {\it not} made
with respect to the models, but relative to photometry of real
galaxies in the Coma cluster.  Systematic zeropoint errors in the high
redshift cluster data seem unlikely, since the data were collected
during many different observing runs with several instruments on
several telescopes.  The infrared Coma photometry we have used to make
the comparison is in good agreement with other published data sets.
Although real evolution at rest frame $\sim$1.1$\mu$m over lookback
times of order 3--5 Gyr seems improbable, we have no other explanation
for the (admittedly small) effect.

There are no dramatic ``outlier'' clusters in our sample for which the
{\it average} properties of their early--type galaxies (colors,
scatter, or $c-m$ slope) are drastically different from those of other
clusters at similar redshift.  This is particularly important given
that the clusters studied here span a broad range of richness and
x--ray luminosity.  Recall that there is no selection criterion for
our sample other than that a cluster has WFPC2 imaging available.
Most of these clusters, therefore, are rich and famous, accounting for
their inclusion in {\it HST} imaging programs, but several
(particularly the radio--selected clusters 3C~295, 3C~220.1 and 3C~34)
are relatively poor in early--type galaxies.  X--ray measurements,
which are available for most of the clusters, show a wide range in
luminosity, from 3.0 $\times 10^{43}$ ergs~s$^{-1}$ for F1557.19TC to
2.0 $\times 10^{45}$ ergs~s$^{-1}$ for MS 0451.6-036 (P.\ Rosati,
personal communication).  The widely--differing optical richnesses of
these two clusters, both of which are at $z \sim 0.5$, correspond to
their substantially different x--ray luminosities.  The average colors
of the early--type populations in these two clusters are nearly the
same, and the same is true for the color scatter.  These similarities
indicate that E+S0 galaxy evolution is independent of cluster
richness, as well as the degree of central concentration.  The
foregoing suggests that early--type galaxy evolution does not depend
strongly on environment, at least above a certain minimum density
threshold corresponding to poor clusters.

\subsection{Color--magnitude Relationship}

We find no evidence for a systematic change of slope in the
color--magnitude relation of early--type cluster galaxies out to $z =
0.9$.  Color evolution is thus essentially independent of luminosity
for cluster early--type galaxies brighter than $\sim$2 magnitude below
$L^\ast$.  This strongly suggests that the slope in the
color--magnitude relation is fundamentally a consequence of
metallicity effects, and not one of age, since less massive cluster
E+S0's show no sign of being younger than their brighter, more massive
counterparts.  There is also no indication of differential chemical
evolution over the lookback time of our sample, which could also
result in a change of slope.  The observed behavior is as expected if
the bulk of star formation in early--type cluster galaxies ceased long
before $z = 0.9$ and the galaxies remained largely fixed in their
original mass--metallicity sequence -- i.e.\ without substantial later
merging to reshuffle the metallicity distributions.

Following the ELS formation scenario, models by Kodama \& Arimoto
(1997) predict little evolution in the color--magnitude slope out to
$z = 1$ if its origin is due to metallicity in stellar populations
formed at early epochs.  However, when those authors match the
present--day color--magnitude slope in clusters with a mass--dependent
age variation among the early--type galaxies, they predict drastic
changes in the slope for $0 < z < 0.9$, which we do not observe.  In
the recent models of Kauffmann \& Charlot (1997), which investigate
elliptical galaxy evolution in the context of hierarchical galaxy
formation with chemical enrichment, the $c-m$ relation is established
and preserved despite extensive merging because more massive and
metal--rich ellipticals are formed by mergers of more massive and
metal--rich spiral progenitors.  In these models, there is a
progressive flattening of the color--magnitude sequence with
increasing redshift.  However, the change in the slope is not dramatic
until $z > 1$.

\subsection{Intrinsic Color Scatter and Early--type Galaxy Formation}

A small scatter in the colors of galaxies at a given luminosity
implies that their stellar populations are similar.  In particular, as
noted previously, various authors (e.g.\ Bower, Lucey, \& Ellis 1992)
have used the small scatter of E+S0 colors in nearby clusters to argue
that the {\it fractional} differences in ages between the stellar
populations of the galaxies are small, setting a joint constraint on
their ages and relative star formation histories.  Either the
early--type galaxies in Coma formed so long ago that color differences
resulting from age variations have mostly damped out, or else their
intrinsic age differences are very small.  If E+S0 galaxies in
clusters share similar (but not identical) star formation histories,
then we would expect the scatter in their colors to increase as one
looks to higher redshifts, i.e. closer to the time when their last
major episodes of star formation took place.

The degree to which the intrinsic color scatter of luminous
early--type galaxies in clusters remains approximately constant over
the very broad lookback time interval of our sample is perhaps the
most striking result of this study.  There is no evidence for a
monotonic trend in the amplitude of the scatter over the redshift
range $0.3 < z < 0.9$ in any of the color combinations we have
examined.  The $blue - K$ and $red - K$ color scatter in the
early--type galaxies for most of the distant clusters do appear to be
systematically larger than is measured in samples of similar galaxies
in the Coma cluster.  While we believe that it may well be real, the
difference is small.  The smaller color scatter among Coma galaxies
may represent a real ``settling down'' of the stellar populations in
early--type cluster galaxies over the past few billion years since $z
= 0.3$, but evidently the scatter was roughly constant over a period
of several Gyr prior to that time.  For our adopted $q_0 = 0.05$
cosmology, the interval of lookback time from $z = 0$ to 0.3 is nearly
the same as that between $z = 0.3$ and 0.9, the redshift range of our
cluster sample; it is slightly longer for $q_0 = 0.5$.  Examining the
scatter in the colors of E+S0 galaxies in clusters at $0.1 < z < 0.3$
should provide an interesting means of studying any transition which
may have taken place during the last few billion years of cluster
evolution.

If the intrinsic color scatter of early--type galaxies in present--day
clusters were due to small age differences among otherwise passively
evolving galaxies, then one would inevitably expect the scatter to
have been larger at earlier times when the galaxies were younger and
their {\it relative} age differences were greater.  Similarly, if
color variation among ellipticals reflects differences in their past
merging histories, then we might expect to see increased scatter as we
look back toward the era when the bulk of this merger activity took
place.  In either case, the resulting color variations will be damped
as the time since the epoch of star formation lengthens, and the
scatter in colors around the $c-m$ relation should diminish toward
$z=0$.  The fact that we do not see such trends over the redshift
interval $0.3 < z < 0.9$ suggests two possibilities.  One is that the
last episodes of major star formation in early--type galaxies took
place several billion years before the lookback times at which we are
observing the clusters, i.e. at redshifts significantly above one.  In
this case the color variations would have time to damp out, leaving a
small and roughly constant value for the observed scatter.
Alternatively, some other effect may act to erase the expected trend
of scatter with redshift predicted from a uniform population.  For
example, some portion of the color scatter in early--type cluster
galaxies could be due to metallicity variations at a fixed galaxy
mass, i.e.\ a scatter in metallicities around the mean trend of the
color--magnitude relation.  This would then set a ``floor'' to the
allowable scatter, which could act to reduce or erase some of the
expected trend with age.  It seems highly plausible that this should
be the case, and such scenarios deserve further investigation,
particularly as multi--metallicity population synthesis models become
more sophisticated.  The models of Kauffmann \& Charlot (1997) produce
cluster ellipticals of a given mass exhibiting a range of
metallicities which partially accounts for the scatter in their
colors.  If, however, the intrinsic scatter is partially a
metallicity effect, then the argument for highly synchronized star
formation histories remains valid, and is perhaps strengthed: the
component of scatter due to age variations must be even smaller than
for models without metallicity variations, requiring a strong degree
of uniformity in the ages of the stellar populations.

Another possibility is that the color scatter at any redshift has 
some component due to more recent episodes of star formation
involving the early--type cluster galaxies.  In this case, the
assumption of purely passive evolution and thus strict coevality
would be incorrect.  The resulting effect on the galaxy colors and 
their scatter would then depend on the rate at which new star formation
occurs.  If the addition of young stars to cluster E+S0s (e.g. during 
merger events) were small, stochastic, and more or less constant with 
redshift, then it could introduce a similar amount of scatter into
the color--magnitude relation over the redshift range we are observing.
To some degree, this interpretation may be related to the Butcher--Oemler
effect in the context of the scenario advocated by Couch \& Sharples
(1987; hereafter CS87) and more recently developed by Barger et al.\
(1995).  In these papers, the Butcher--Oemler effect is described as a 
starburst phase through which nearly {\it all} cluster galaxies pass at some 
point in their lifetimes.  Evidence in favor of such a scenario includes 
the red $H\delta$--strong galaxies of CS87 and the E+A galaxies of
Dressler \& Gunn (1983), though the latter may no longer be supportive
given that WFPC2 imaging has shown that the E+A galaxies are generally
disks (Wirth, Koo, \& Kron 1994; Belloni et al.\ 1996).  In the CS87
scenario, small amounts of recent star formation in the early--type
galaxies, which might go undetected in terms of the effect on the
average colors seen in Figure 3, would show up as increased color
scatter.  However, the monotonic color evolution trend with redshift 
seen in Figure 3 argues that such events are relatively unimportant 
to the overall stellar population.

The small scatter found in the early--type galaxy colors at the
highest redshift in our sample places the most stringent constraint on
the formation redshift and/or the coevality of the early--type galaxy
populations.  Following Bower, Lucey, \& Ellis (1992), a simple
comparison of the rate of change in the colors of a single--burst
stellar population (BC96) model to the estimated intrinsic scatter
yields a lower limit to the galaxy age at the cluster redshift.  The
intrinsic scatter in the observed frame colors $blue-K$ and $red-K$ of
the early--types in GHO~1603+4313 is $0.11\pm0.06$ and $0.10\pm0.04$,
respectively.  If early--type galaxy formation within this cluster
occurred at times distributed uniformly across the collapse epoch
(i.e.\ the $\beta = 1$ case of Bower et al.), then the mean galaxy age
at $z=0.895$ must be at least 3.9 Gyr.  The resulting limit to the
formation redshift is $z \ge 3.0 ^{+0.0}_{-0.3}$.  This limit is
consistent with the acceptable BC96 models for color evolution
described in \S 4.1.  If galaxy formation occurred later, then
the degree of synchronization must have been correspondingly greater:
ages of $\sim$2~Gyr are admissible if galaxy formation were
synchronized within a time interval of 500 million years.  Again
following the lead of Bower et al., the method described above can be
turned around to estimate the amount of recent star formation.
Assuming a fundamentally ``old,'' passively evolving model for
early--type galaxy evolution, the intrinsic color scatter in
GHO~1603+4313 limits small (e.g.\ $\le$10 \% by mass) bursts to have
occurred at least 2 Gyr prior to its cosmic age at $z = 0.9$.

To summarize, the simplest interpretation of the scatter trends is that 
the stellar populations in most cluster early--type galaxies have been
quiescent since at least $z \approx 1.5$.  This ensemble behavior does
not exclude the possibility that {\it some} cluster ellipticals or S0s
may have experienced star formation since that time.  However, the striking 
similarity between the $c-m$ relations for early--type galaxies in clusters
at redshifts as large as 0.9 (e.g. Figure~2) and those in Coma today certainly
suggests a strong degree of evolutionary continuity.  Altogether, it remains 
to be seen whether realistic cluster and cluster galaxy evolution models 
can be constructed which match both the Butcher--Oemler effect in disk 
galaxies and the flat behavior of the color scatter vs.\ redshift for 
early--type galaxies.

\subsection{Selection Effects}

While the quiescent photometric behavior of early--type galaxies in
our cluster sample strongly suggests a quiescent evolutionary history
out to $z \approx 1$, it is important to consider the potential role of
the sample selection criteria in predetermining such behavior.  There
are three primary selection criteria for galaxies in our sample:
location in the cores of rich clusters; E or S0 morphology; and
luminosity in the observed $K$ band.  As noted in the introduction,
selection by rest frame near-IR luminosity should yield a mix of
galaxy types which does not strongly depend on redshift, alleviating
the bias towards UV bright galaxies at higher redshift.  Therefore
while one would expect more quiescent behavior in a $K$--selected
sample than in an optically selected sample, $K$--selected samples
provide a truer picture of intrinsic changes with redshift.  However,
the other selection criteria may be less benign.

Morphological galaxy selection may introduce a type of bias.
Within each cluster, WFPC2 imaging has provided a sample of galaxies 
recognizable as morphological early--types.  However, if an elliptical 
or S0 galaxy were undergoing an interaction or were forming during a 
major merger event, it might not be classified as an ``early type'' and 
thus could be excluded from our sample.  By the time the merger
morphology has settled down sufficiently for the galaxy to be classified
as a normal E or S0 (i.e.\ when tidal tails, star forming knots, multiple 
nuclei, remnant spiral arms, etc. have fully merged or faded away),
much of the blue light from recent star formation may also have subsided,
leaving only the dominant old stellar light.  In this way, sample
definition might partially account for the small scatter in galaxy
colors which we observe.  If the antecendents of some of today's 
early--type galaxies cannot be recognized by their morphologies at
high redshift, then it will be difficult to wisely choose adequate 
samples for comparison with present--epoch cluster ellipticals.  
The use of WFPC2 imaging to trace the evolution of galaxies divided
according to their Hubble types becomes more complicated if 
morphological classes among galaxies are not conserved.
An investigation of the morphological effects of interactions on
galaxies in high redshift clusters, which is beyond the
scope of our study but has been explored through simulations by
e.g. Moore et al.\ 1996, would be necessary to quantify this 
effect.

Another potential problem with our morphological selection is the lack
of distinction we make between E and S0 morphologies.  Though
apparently similar today in many ways, elliptical and S0 galaxies may
also have important differences.  If they have followed different
evolutionary histories, this may affect conclusions reached from an
analysis of their joint photometric properties.

Dressler et al.\ (1997), using WFPC2 imaging data, have reported 
evidence for an increase in the proportion of S0 galaxies in clusters 
from $z \sim 0.5$ to the present (see also Smail et al.\ 1997).
Recall that we have not attempted to distinguish between elliptical 
and S0 galaxies in our samples.  If the Butcher--Oemler effect were due 
to disk galaxies evolving to become S0s between $z \sim 0.5$ and $z=0$ 
then this may have some effect on the galaxy colors and color scatter 
which we observe.  On the one hand, if the transformation from spiral 
to lenticular is accompanied by star formation, or is the result of 
the truncation of star formation in the pre--existing spiral galaxy 
disks, then ``younger'' S0s at higher redshifts should be bluer, 
affecting both their mean colors and the scatter in those colors.
However, at some point, the S0 progenitors would no longer be classified
as S0s on the basis of their morphologies, but as later--type disk galaxies
instead, and thus would ``drop out'' of our early--type galaxy sample 
at larger redshifts.  The influence of their bluer colors on the mean
photometric trends we are studying here would then cease.

We can assess some of these issues in a preliminary way by looking at
the fraction of E+S0 galaxies in the cores of clusters vs. redshift.
The numbers listed in Table 1 can be used to estimate the ratio of
field--corrected early type cluster galaxies to field--corrected
cluster galaxies of all types within three redshift bins: $0.3 < z <
0.45$, $0.45 < z < 0.65$, and $0.65 < z < 0.9$.  Averaging with
uniform weight for each cluster, the fraction of early--type to total
cluster galaxies is 0.64, 0.78, and 0.69 respectively for these
redshift bins.  Averaging with uniform weight for each cluster {\it
galaxy} within a redshift bin the fractions are 0.61, 0.74, and 0.8.
This crude analysis does not reveal the decline in the frequency of
early--type galaxies in high redshift clusters which might be expected
if E+S0 galaxies form from mergers of later type galaxies, or S0's in
$z \sim 0.3$ clusters form from later type spirals at higher
redshift.  However, it is worth noting that the fraction of
early--type galaxies in the Coma-cluster sample used for reference
here is 0.96, again hinting at a transition in cluster galaxy
properties between $0 < z < 0.3$.

There is indeed evidence for stellar population differences between 
elliptical and lenticular galaxies in the local universe.  Bothun \& Gregg
(1990) have used optical--IR photometry of the bulge and disk components
of S0 galaxies to show that the disks appear to be younger (more intermediate
age stars) than the bulges.  Schweizer \& Seitzer (1992) examined the
$UBV$ color--magnitude relation for E and S0 galaxies separately, and found
that while the mean slopes and intercepts of the relations were the same
(see also Sandage \& Visvanathan 1978), the color scatter about the mean 
is substantially larger for the S0s.  However, it is worth noting that the 
observational samples for both of these studies consisted primarily of 
{\it field} galaxies, and not the inhabitants of the cores of rich clusters.
Considering ellipticals and S0s separately in the Coma cluster data used
here for our $z=0$ reference sample, De~Propris et al.\ (1997) find little
indication of significantly larger scatter in the S0 colors.  Nevertheless,
it is important to keep in mind that elliptical and S0 galaxies may have
followed different evolutionary paths.  It would be very interesting 
to repeat our analysis for samples of distant cluster ellipticals and 
S0s separately to search for different behavior.  This might be accomplished 
using new, precise morphological catalogs (e.g. Smail et al.\ 1997).  
However, the accurate separation of S0s from ellipticals at $z > 0.5$ 
using WFPC2 imaging still needs to be carefully investigated.

Even within the traditional, passive evolution scenarios, it is
important to consider possible selection effects resulting from the
choice of early--type galaxies in clusters as a sample to investigate.
Galaxy clusters are rare objects, far out on the high mass tail of
bound structures in the universe.  In hierarchical models with $\Omega
= 1$, this is increasingly true at higher redshifts --- massive, bound
clusters should be increasingly unusual objects, formed from larger
and more extreme peaks in the primordial spectrum of mass
fluctuations.  Kauffmann (1996) has appealed to this as part of the
explanation of the Butcher--Oemler effect, and it may play a role in
our understanding of the early--type cluster galaxies as well.  In
biased galaxy formation scenarios, galaxies in the vicinity of
large--scale overdensities may collapse and form earlier than those in
less dense regions.  It may be that by studying only clusters, we
would be increasingly biased toward ``old'' galaxies at higher
redshifts.  Indeed, according to Kauffmann \& Charlot (1997), this is
the reason heirarchical merging models for rich clusters display
elliptical galaxy evolution which appears indistinguishable from that
predicted by the classical passive scenario.  In these models, rich
clusters at high redshift formed earlier than clusters with comparable
circular velocities today, and correspondingly the star formation and
merging histories of their galaxies are pushed back to earlier times.
By modeling (or observing) apparently similar clusters at each
redshift, one does not trace the continuous evolution of a single
collection of galaxies, but instead selects only the oldest galaxies
at any redshift, which thus mimic passive evolutionary behavior.

A natural consequence of such models would be that the evolution of
elliptical galaxies should depend on their environments.   Evidence
for intermediate age stellar populations in some elliptical galaxies
today (e.g.\ Worthey 1996), which may well correlate with their
environment (Bower et al.\ 1990), would support this idea.   Arguing
against this, however, is the fact that our distant cluster sample does
in fact span a broad range of optical richness and x--ray luminosity,
and that the observed evolutionary trends seem to be independent of these
environmental factors.  It is also worth stressing that the hierarchical
models of (e.g.) Kauffman \& Charlot (1997) consider only $\Omega = 1$
CDM universes.  In an open universe, structure formation progresses
differently, with rich clusters forming earlier and ``freezing out''
at comparatively high redshift.   Presumably, the history of cluster
ellipticals in an open universe would also proceed differently, and 
might more closely resemble classical passive evolution across a broader
range of environments.  Extending the observations presented here,
as well as the theoretical models, to early--type galaxies in high redshift
groups and in the field would add another useful dimension to the problem.

\section{Conclusions}

The color evolution seen in Fig.\ 3 is consistent with the idea that
luminous early--type cluster galaxies have evolved with time, and that
their stellar populations were younger at high redshifts.  In itself
this is no surprise; it is the uniformity and regularity of the 
spectrophotometric evolution which is the more important result.  The
degree of color evolution which we find is mild.  At $z = 0.8$ to 0.9,
the observed $R-K$ colors correspond approximately to rest--frame $U-J$.
Early--type cluster galaxies at that epoch were only 0.4 to 0.6 magnitudes 
bluer at these wavelengths than are galaxies of the same morphological 
types in the Coma cluster today.  Most of this change in color occurs 
because of an increase in the near--UV flux --- the IR--IR colors change 
very little.  The mean ages of the stars in E+S0 galaxies in the distant 
clusters are thus younger than those in their present day counterparts, 
but their intrinsic colors are still quite red out to nearly $z=1$.
Spectrophotometric models of passive evolution predict that UV--to--IR
color evolution should be rapid during the first $\sim$2 Gyrs after
star formation ceases, quickly reaching red colors which evolve in a
slow and steady fashion thereafter.  Looking back to $z = 0.9$ (5.6
Gyr ago for our adopted cosmology), we evidently have not yet closely
approached the time at which the bulk of the stars in early--type
cluster galaxies were young.

This conclusion is reinforced by the consistently small scatter in the
galaxy colors out to $z = 0.9$.  The rms dispersion in the $blue - K$
colors remains fixed at approximately 0.1 magnitudes over the redshift
range of our sample, without evidence for the substantial change that
might be expected if the early--type cluster galaxies formed over a
broad range of redshifts.  It is reasonable to expect some component
of the intrinsic color scatter at all redshifts is due to metallicity
variations, and that overall the evolution of early--type galaxies
within a given cluster is highly synchronized.  In addition, we see
little evidence of substantial variations in the average galaxy colors 
from cluster to cluster at a fixed redshift, despite the fact that 
the clusters span a wide range of richnesses and x--ray luminosities.  
The constancy with redshift of the slope in the color--magnitude relation 
of the luminous E+S0 cluster galaxies strongly favors a scenario in 
which this slope arises from a mass--metallicity correlation, rather than 
one involving age.  It appears that, on average, the evolutionary history of 
the majority of luminous early--type cluster galaxies has been very similar.

It is important to recall that the redshift dependence of the galaxy
colors shown in Figure 3 describes the evolution of the {\it stellar
populations} within those galaxies rather than the evolution of the
{\it galaxies} themselves.  The histories of galaxies and of the stars
from which they are made may, in principle, differ substantially.  It
is for this reason that hierarchical merging models, as described in
the introduction, are able to match the color properties of nearby
cluster galaxies, despite the fact that substantial merging takes place
at relatively late times.  In those models, the bulk of the {\it stars}
which comprise today's elliptical galaxies form early --- the assembly
of the galaxy itself occurs fairly late.  However, this merging is
unlikely to occur without some impact on the galaxy colors due to
episodes of star formation during the merging process.  In particular,
it has often been noted that mergers of disk galaxies cannot produce
the high phase--space density of stars observed in the central regions
of ellipticals unless some amount of star formation takes place in gas
funneled to the nucleus of the merger remnant  (e.g. Carlberg 1986;
Hernquist, Spergel \& Heyl 1993; Barnes \& Hernquist 1996).  If most
elliptical galaxies in clusters form by a process of late mergers,
there should be some detectable signature on their overall photometric
properties such as the scatter in their colors, or the slope of the
color--magnitude relation.  By extending the redshift baseline for
cluster photometric studies to $z = 0.9$, a redshift at which
hierarchical models predict that galaxy formation is only partially
complete, we have provided a data set against which detailed models can
be tested.  It seems that it is not difficult to reconcile the observed
color evolution with traditional models of monolithic collapse and
passive evolution for early--type galaxies, but it remains to be seen
whether other scenarios can account for the observations.  In this
context, it will be important to examine the luminosity functions of
cluster galaxies as a function of redshift and of morphological type,
as these should be more sensitive to the merging history of the
galaxies than the colors themselves.

The upper redshift limit of our cluster sample just reaches the point
at which serious constraints can be placed on the evolutionary history
of early--type galaxies.  Beyond $z\sim1$ in cosmologically flat CDM
models, the amount of merging occurring within the prior $\sim$1 Gyr
is sufficiently large so as to seriously inflate the locus of
early--type galaxy colors, even if the recent, merger--induced
starbursts are small (Kauffmann 1996).  The identification of clusters
at $z > 1$ and the characterization of their galaxy populations should
provide a powerful means of testing galaxy formation theories.

\acknowledgments

The authors would like to thank NOAO for a generous allocation of
observing time to this project, and the staffs at Kitt Peak and Cerro
Tololo for their help with the observing.  The referee helped us to
improve the final version by providing a valuable critique of the
submitted version of this paper.  We also thank Marcia Rieke for
allowing us to use her unpublished elliptical spectrum, Megan Donahue
for allowing us to use her WFPC2 image of MS1054.5-0321, and the many
investigators who originally collected the {\it HST} data used in our
archival program.  Stephane Charlot has provided much advice and
guidance concerning the population synthesis models, as well as many
interesting discussions about cluster galaxy evolution.  Support for
this work was provided by NASA through grant number AR--5790.02--94A
from the Space Telescope Science Institute, which is operated by the
Association of Universities for Research in Astronomy, Inc., under
NASA contract NAS5-26555.  Portions of the research described here
were carried out by the Jet Propulsion Laboratory, California
Institute of Technology, under a contract with NASA.  Work performed
at the Lawrence Livermore National Laboratory is supported by the DOE
under contract W7405-ENG-48.

\newpage

\begin{figure}

\figurenum{1}
\plotone{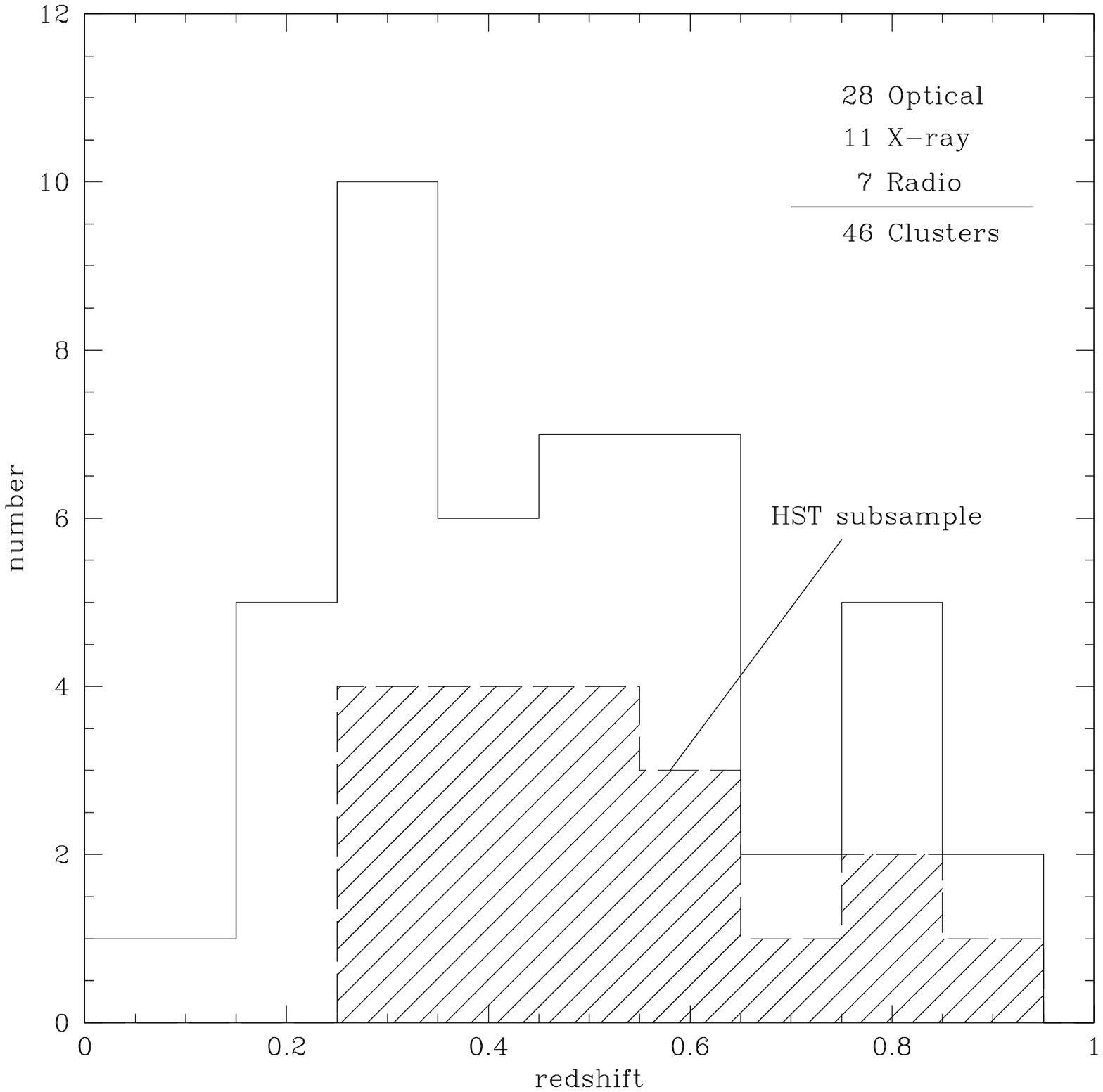}
\caption{Redshift distribution of clusters in our complete imaging survey.
Clusters with {\it HST} imaging, which are the subject of this paper, 
are shaded.}

\end{figure}

\begin{figure}
\figurenum{2}
\epsscale{0.8}
\plotone{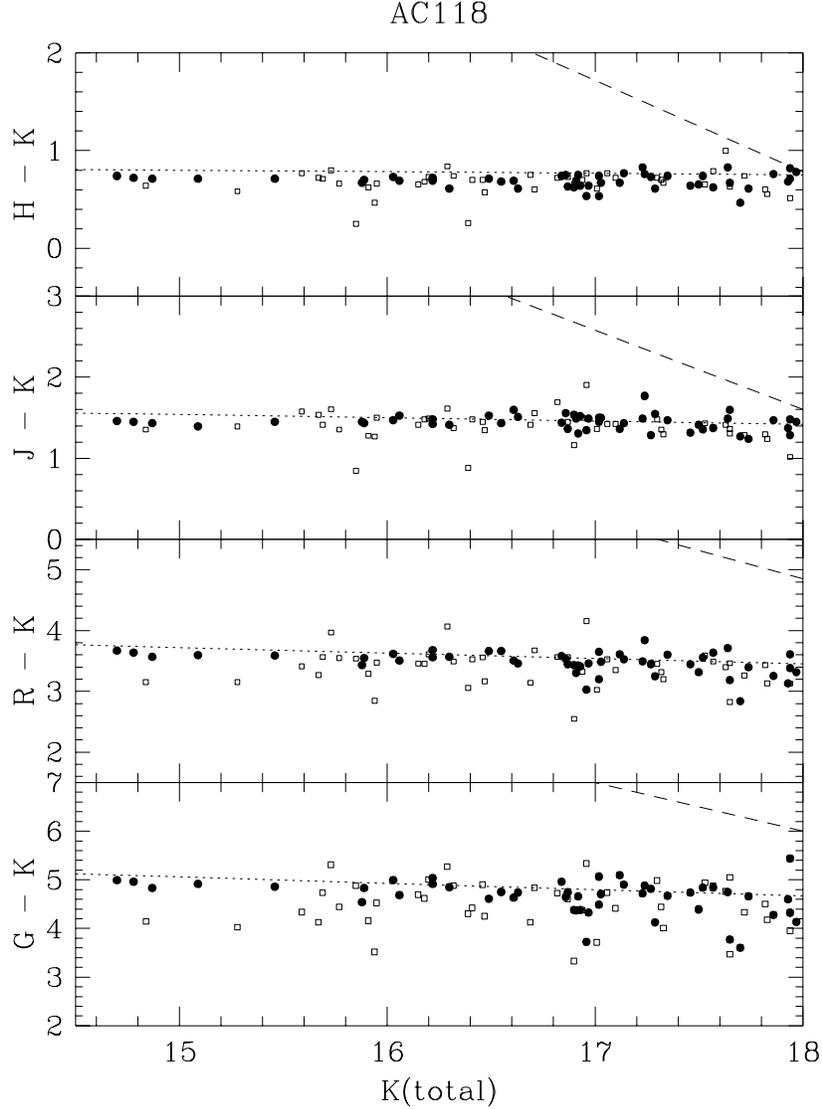}
\caption{(a--q) Color--magnitude diagrams for all of the clusters
in the $HST$ subsample, except for Abells 370 and 851 (previously
published in SED95).  The clusters are arranged in redshift order, from
low to high.  The filled circles are the morphologically--selected
early--type galaxies in the WFPC2 field, and the open squares are all
the other objects in the same area.  Morphological identification was
carried out for all objects in the WFPC2 field down to a limiting $K$
magnitude of $K^\ast + 2$ mag (except for 3C 34; see text for
details).  The slanting dashed lines mark the 5 $\sigma$ limits in
each color.  The dotted lines show the color--magnitude relation for
E+S0 galaxies in the Coma cluster transformed out to the redshift of
each cluster as described in the text; these represent the
no--evolution loci.}
\end{figure}
\clearpage
\begin{figure}
\figurenum{2b}
\epsscale{0.9}
\plotone{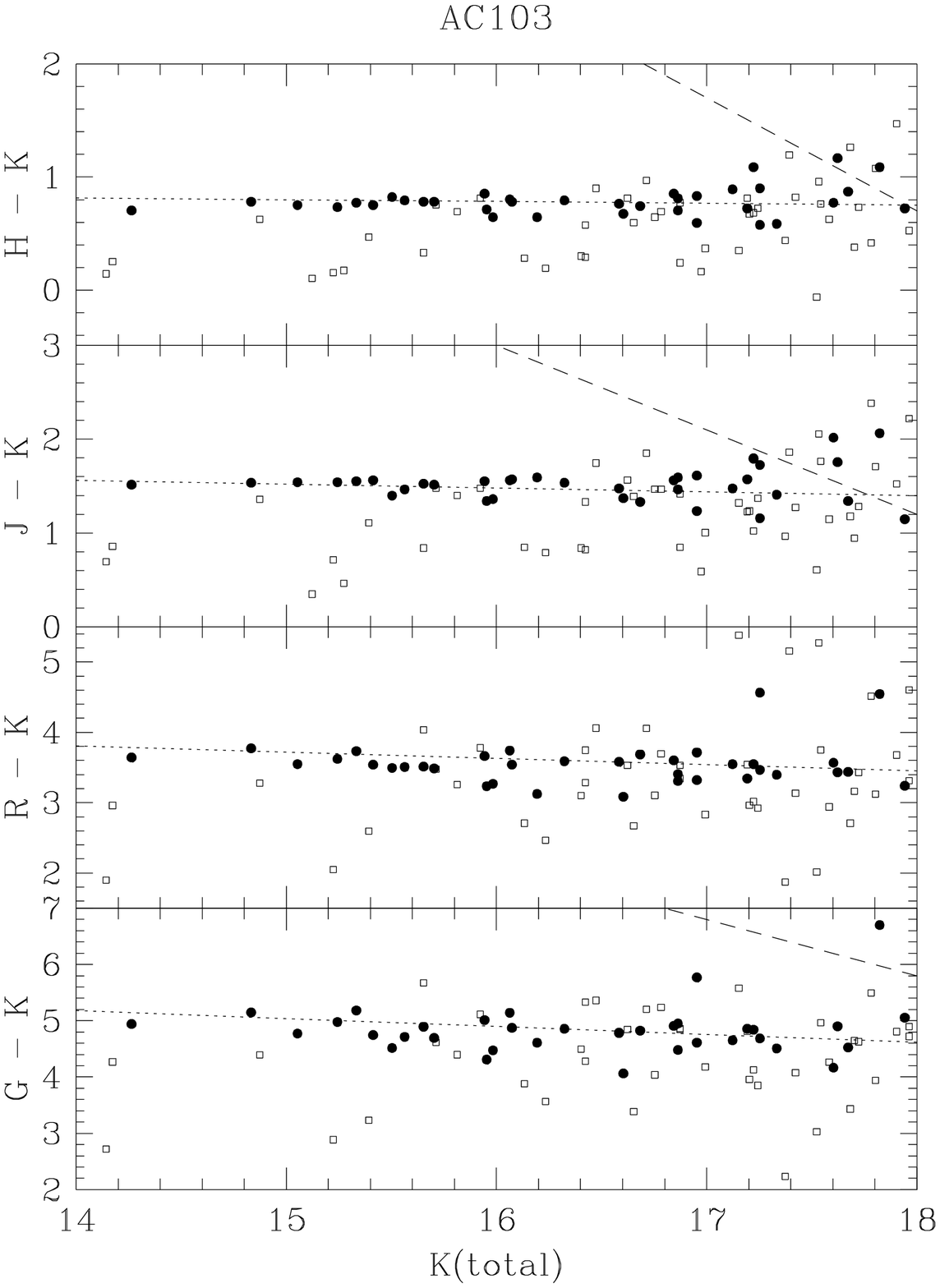}
\caption{}
\end{figure}
\clearpage
\begin{figure}
\figurenum{2c}
\epsscale{0.9}
\plotone{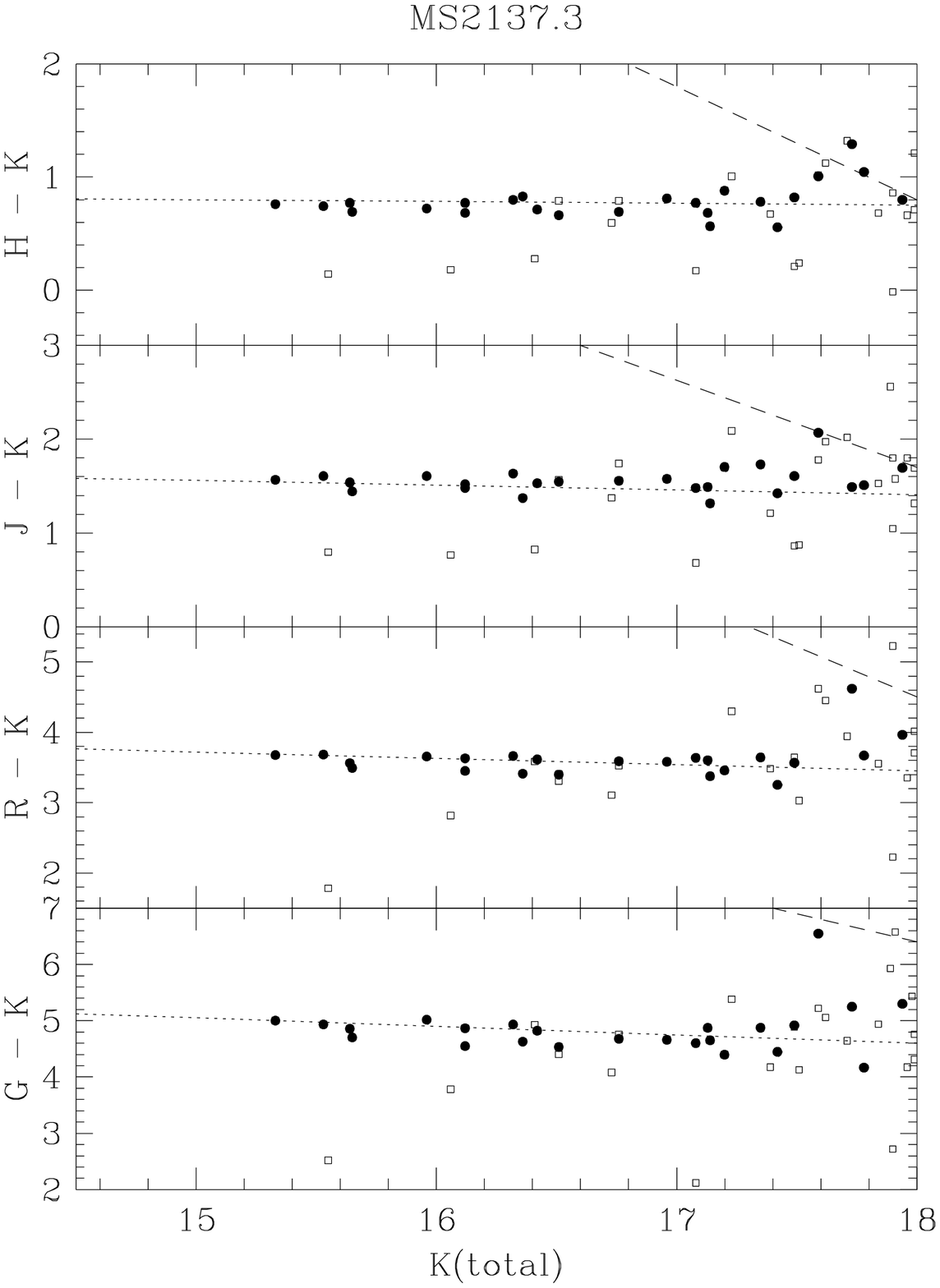}
\caption{}
\end{figure}
\clearpage
\begin{figure}
\figurenum{2d}
\epsscale{0.9}
\plotone{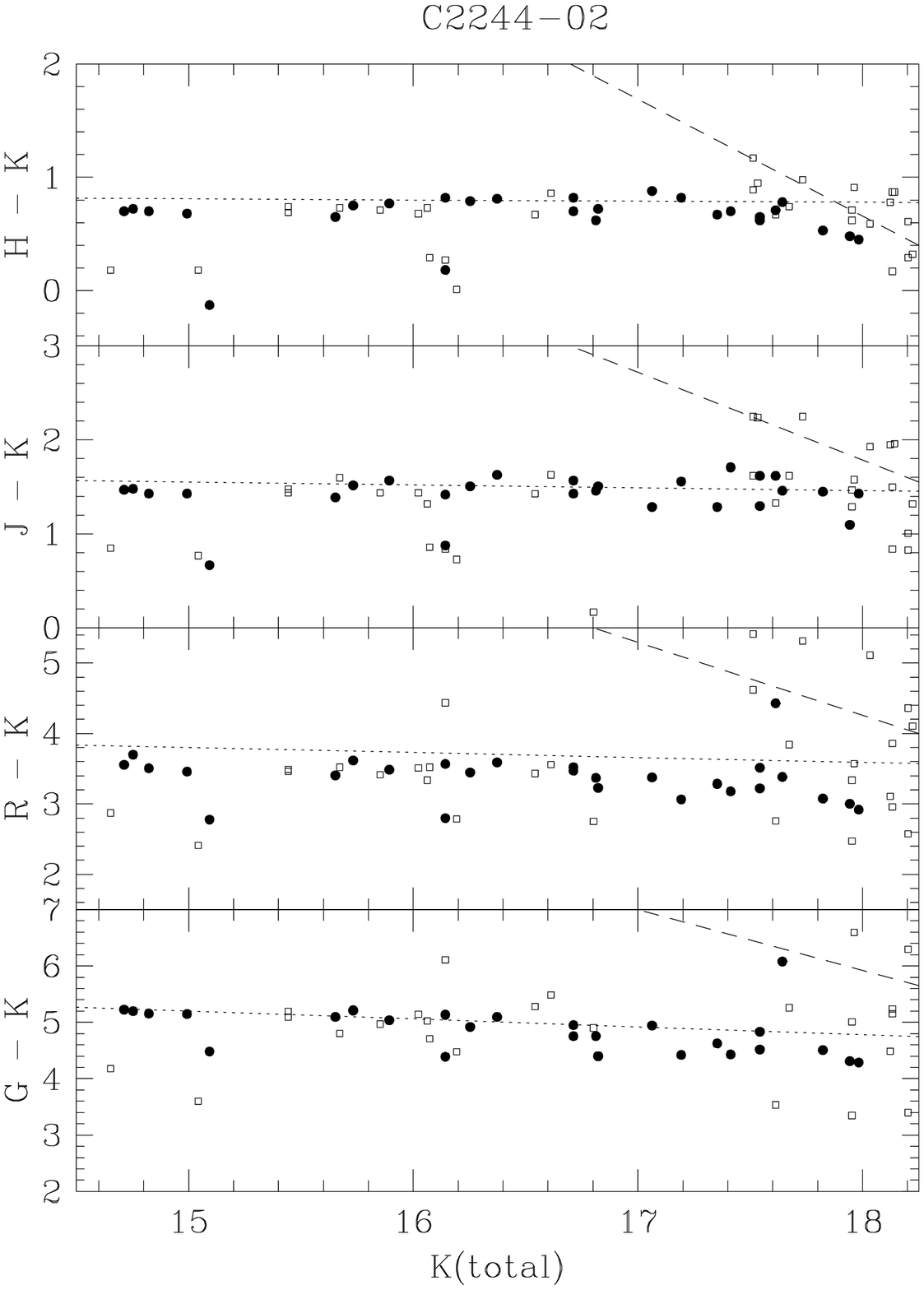}
\caption{}
\end{figure}
\clearpage
\begin{figure}
\figurenum{2e}
\epsscale{0.9}
\plotone{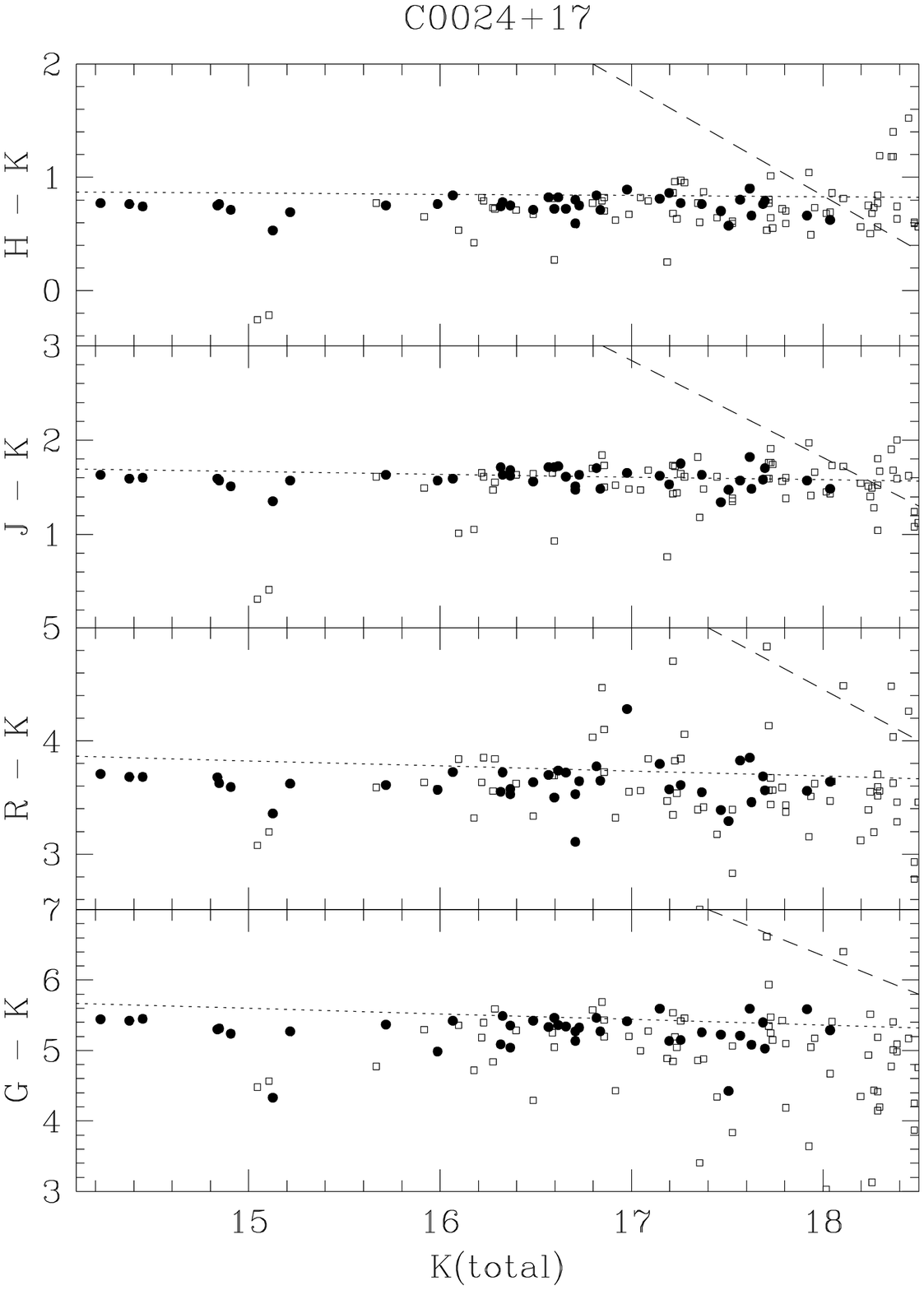}
\caption{}
\end{figure}
\clearpage
\begin{figure}
\figurenum{2f}
\epsscale{0.9}
\plotone{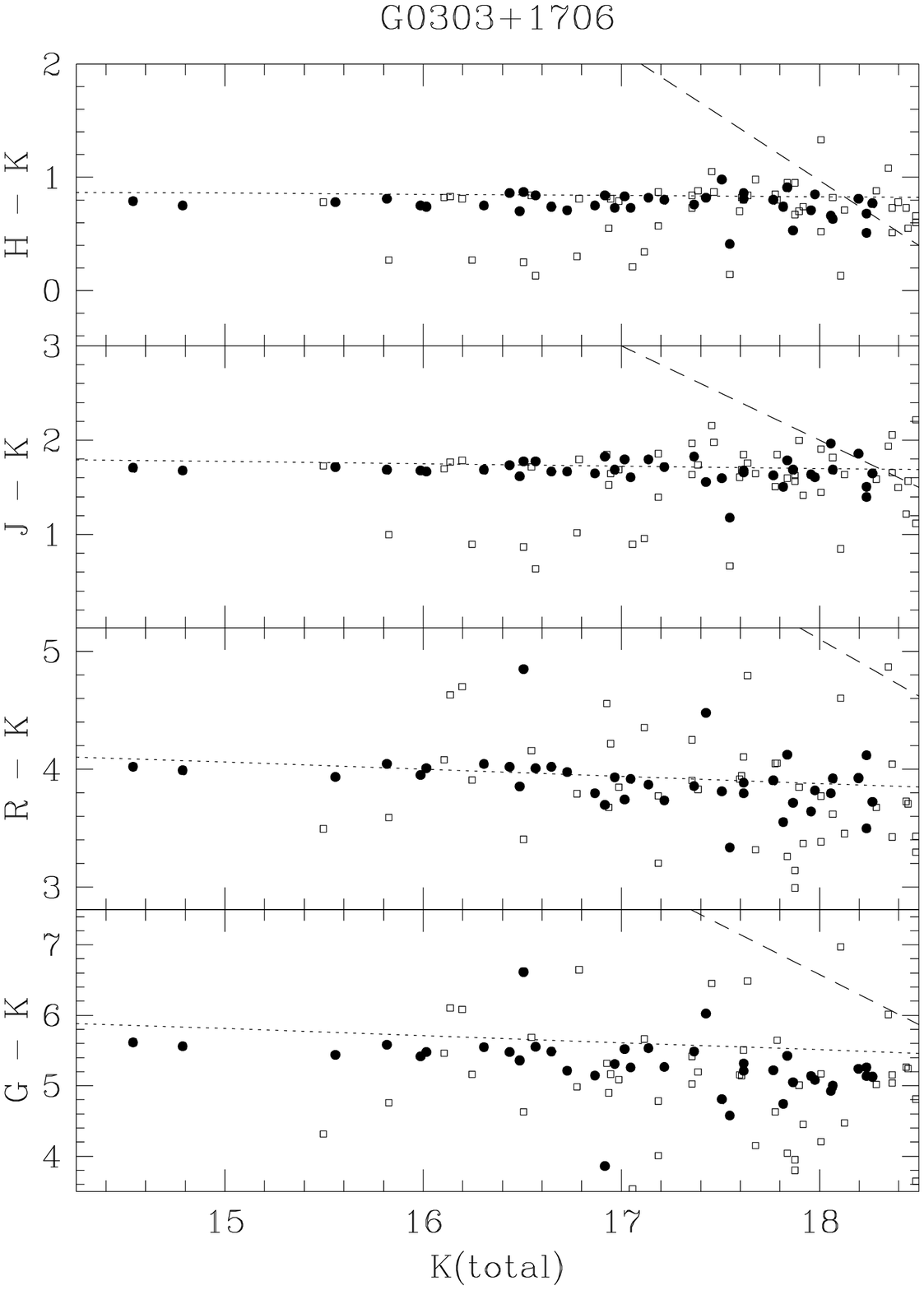}
\caption{}
\end{figure}
\clearpage
\begin{figure}
\figurenum{2g}
\epsscale{0.9}
\plotone{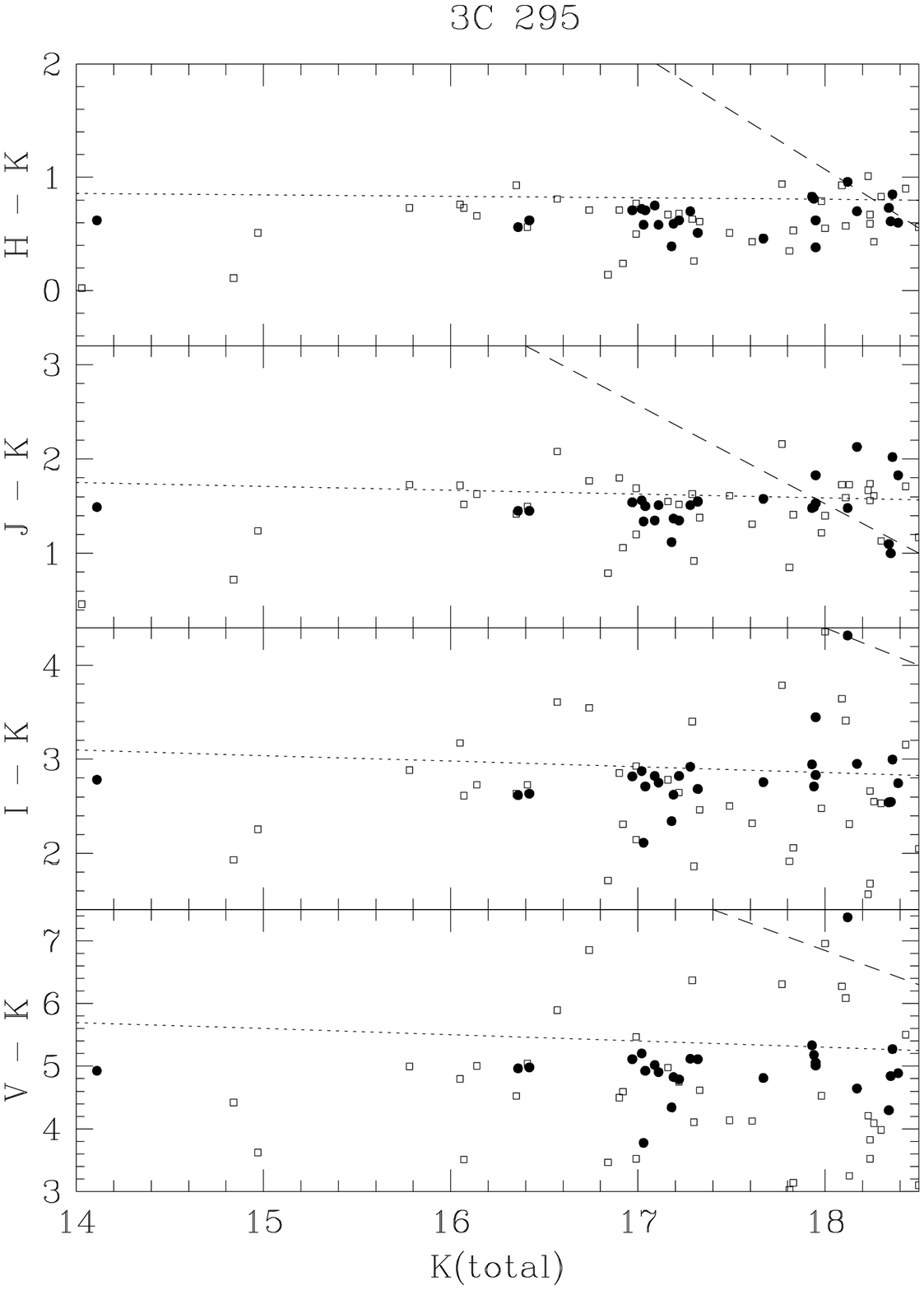}
\caption{}
\end{figure}
\clearpage
\begin{figure}
\figurenum{2h}
\epsscale{0.9}
\plotone{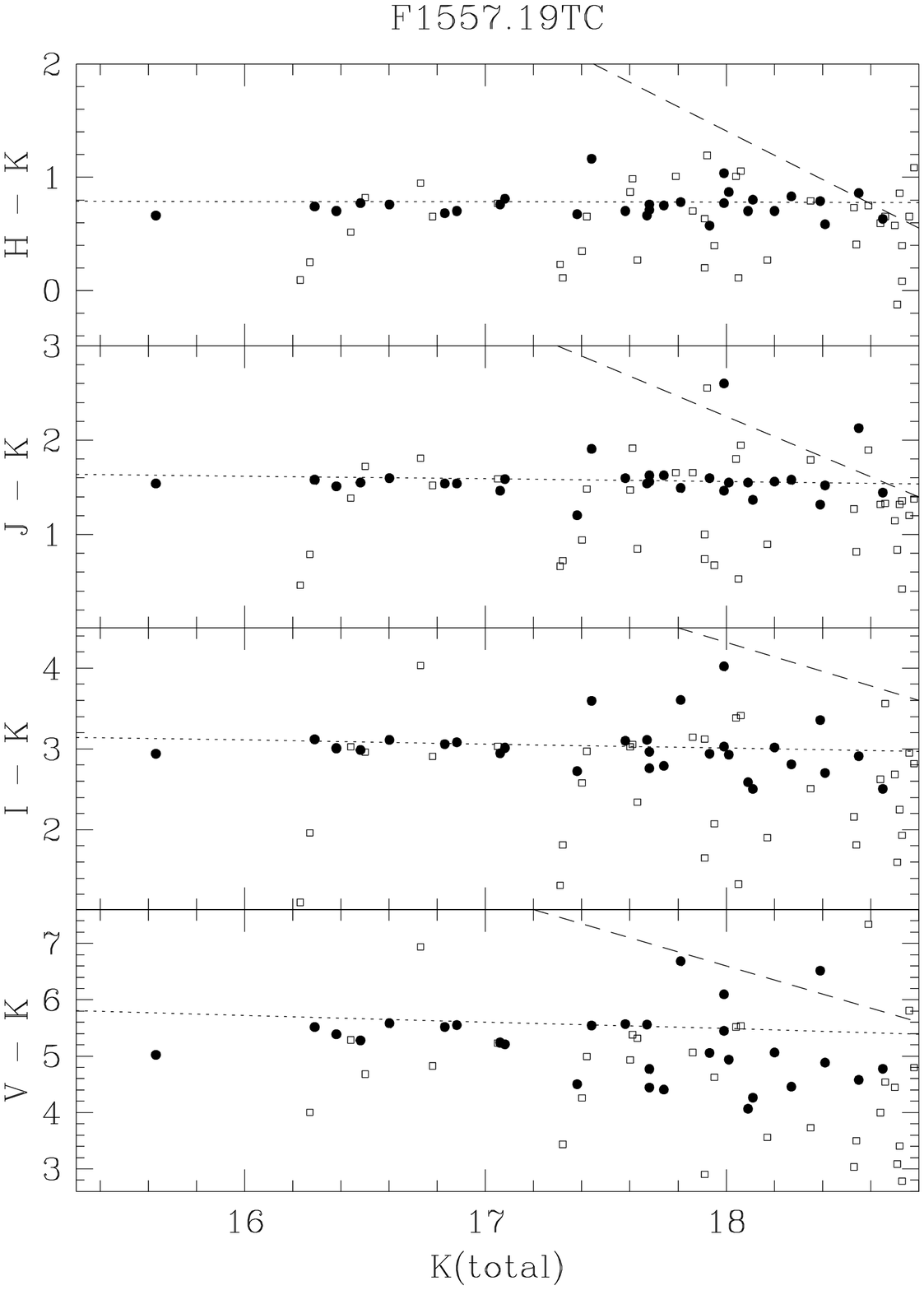}
\caption{}
\end{figure}
\clearpage
\begin{figure}
\figurenum{2i}
\epsscale{0.9}
\plotone{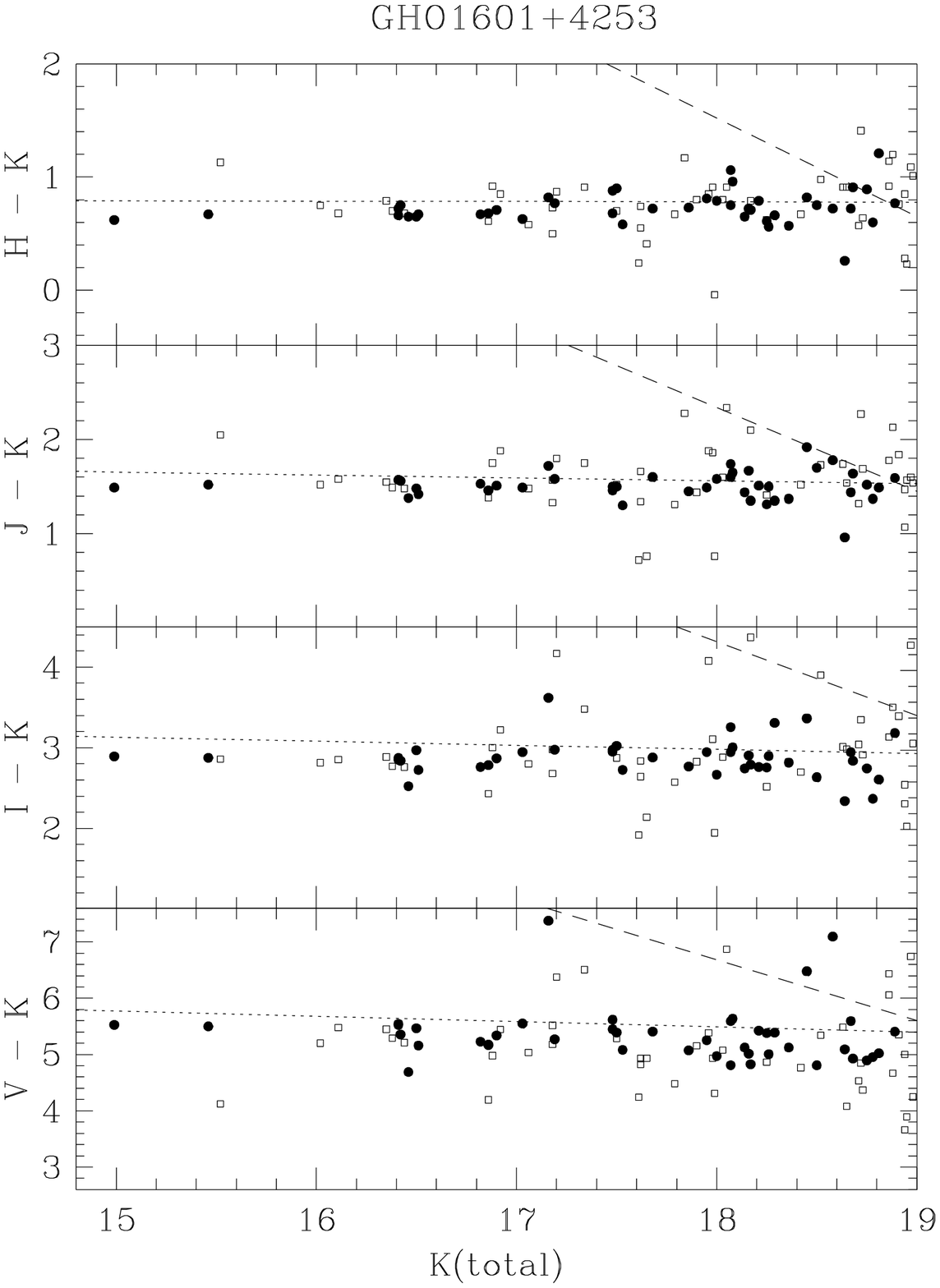}
\caption{}
\end{figure}
\clearpage
\begin{figure}
\figurenum{2j}
\epsscale{0.9}
\plotone{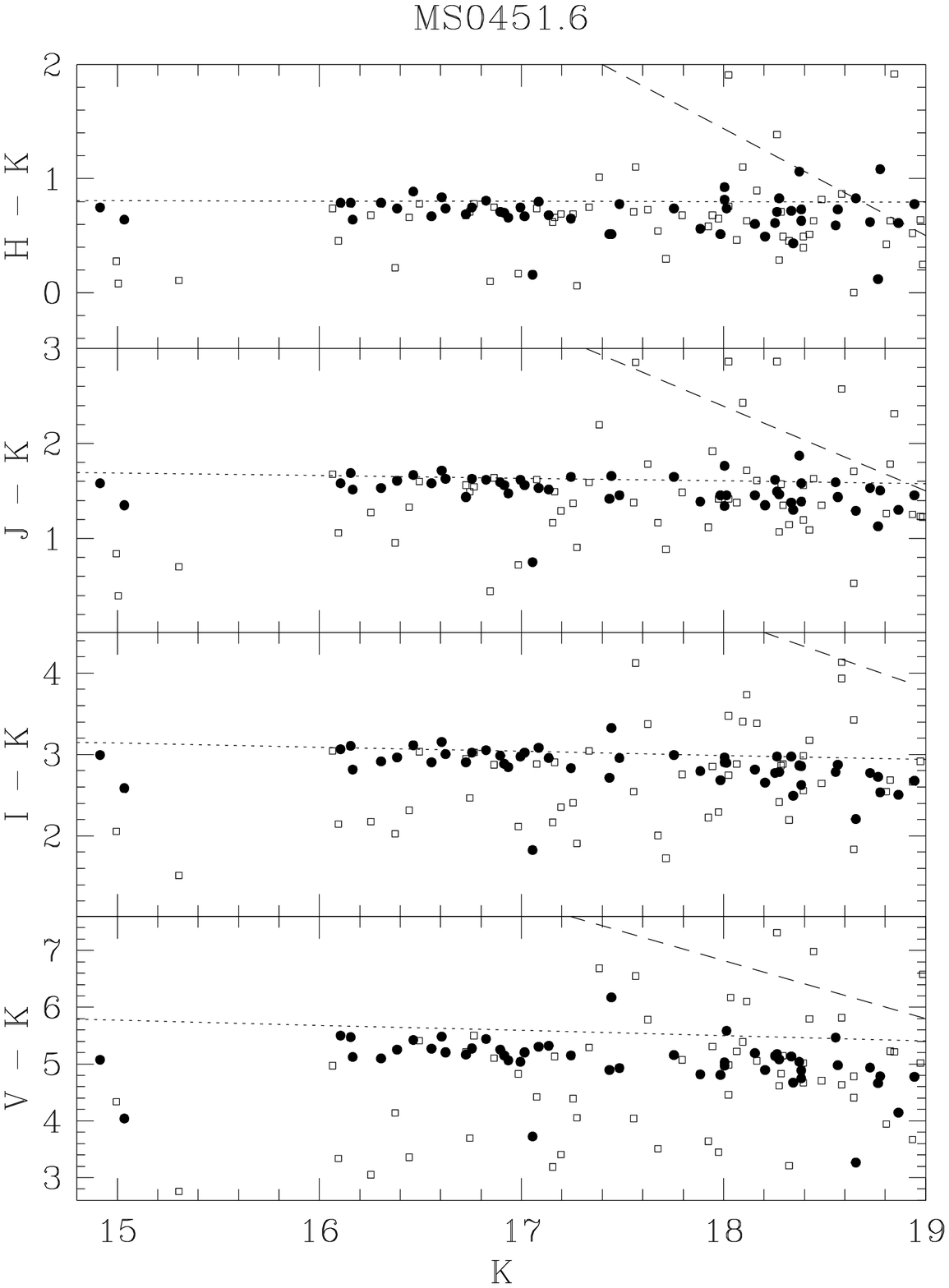}
\caption{}
\end{figure}
\clearpage
\begin{figure}
\figurenum{2k}
\epsscale{0.9}
\plotone{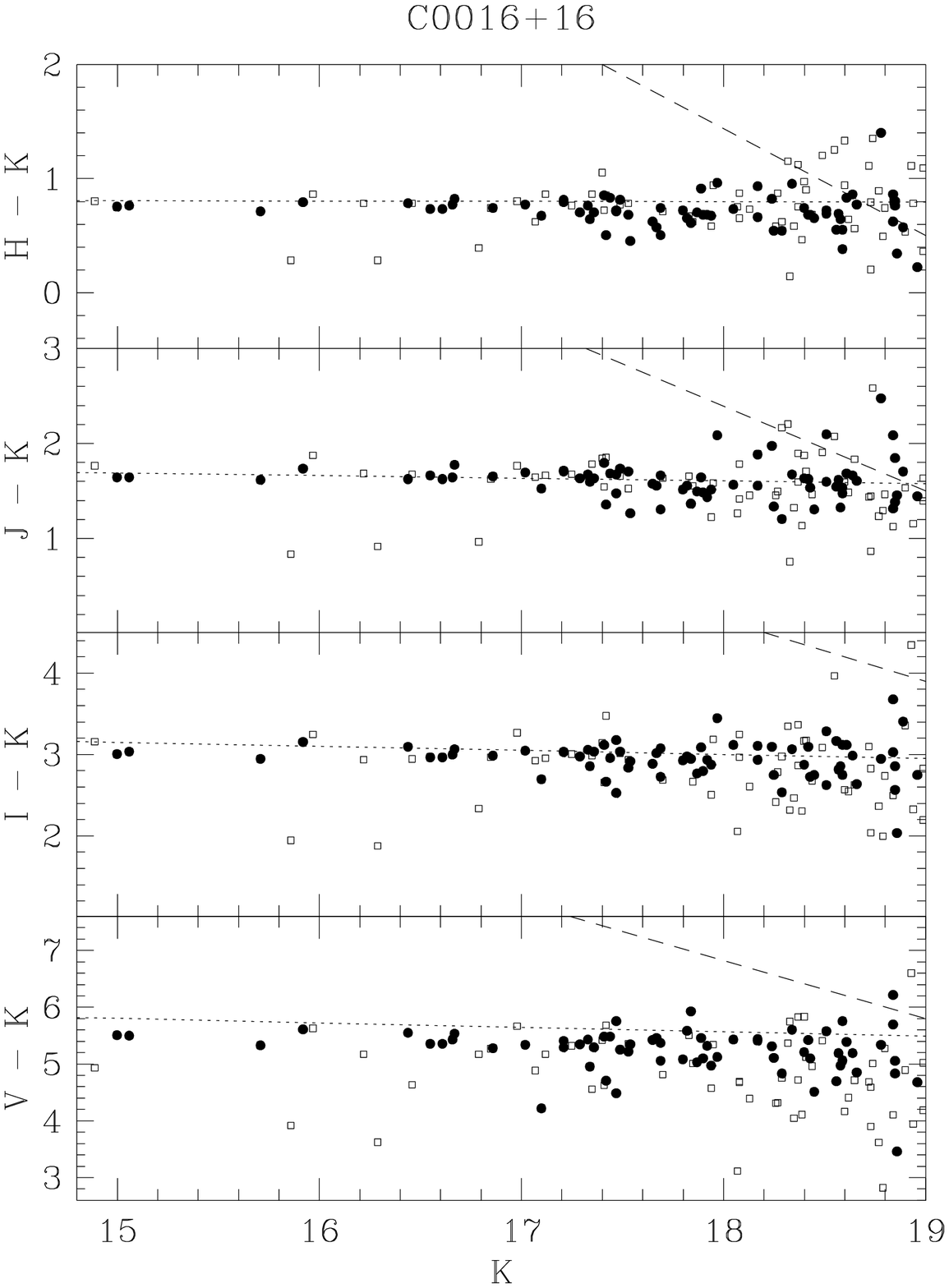}
\caption{}
\end{figure}
\clearpage
\begin{figure}
\figurenum{2l}
\epsscale{0.9}
\plotone{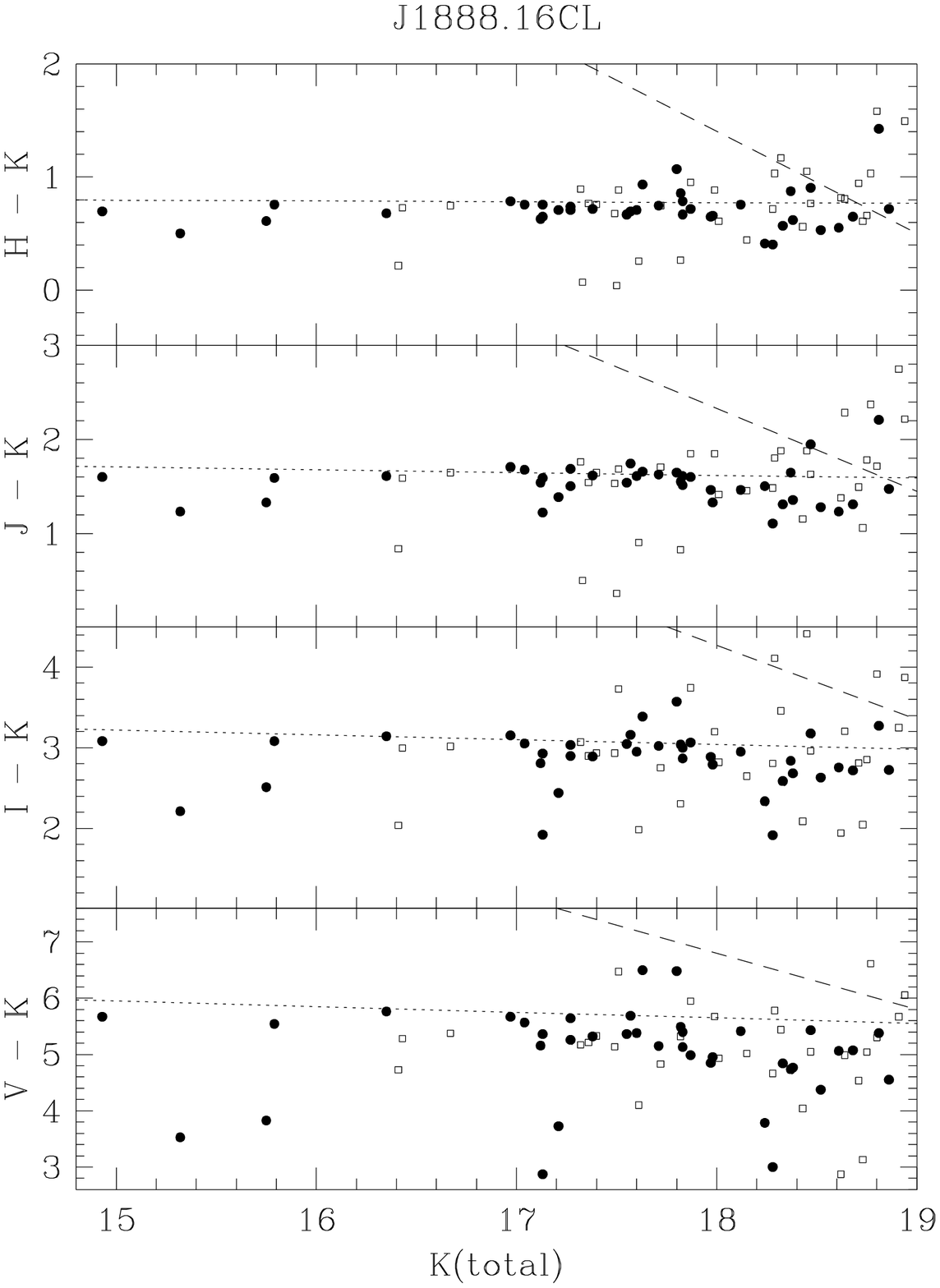}
\caption{}
\end{figure}
\clearpage
\begin{figure}
\figurenum{2m}
\epsscale{0.9}
\plotone{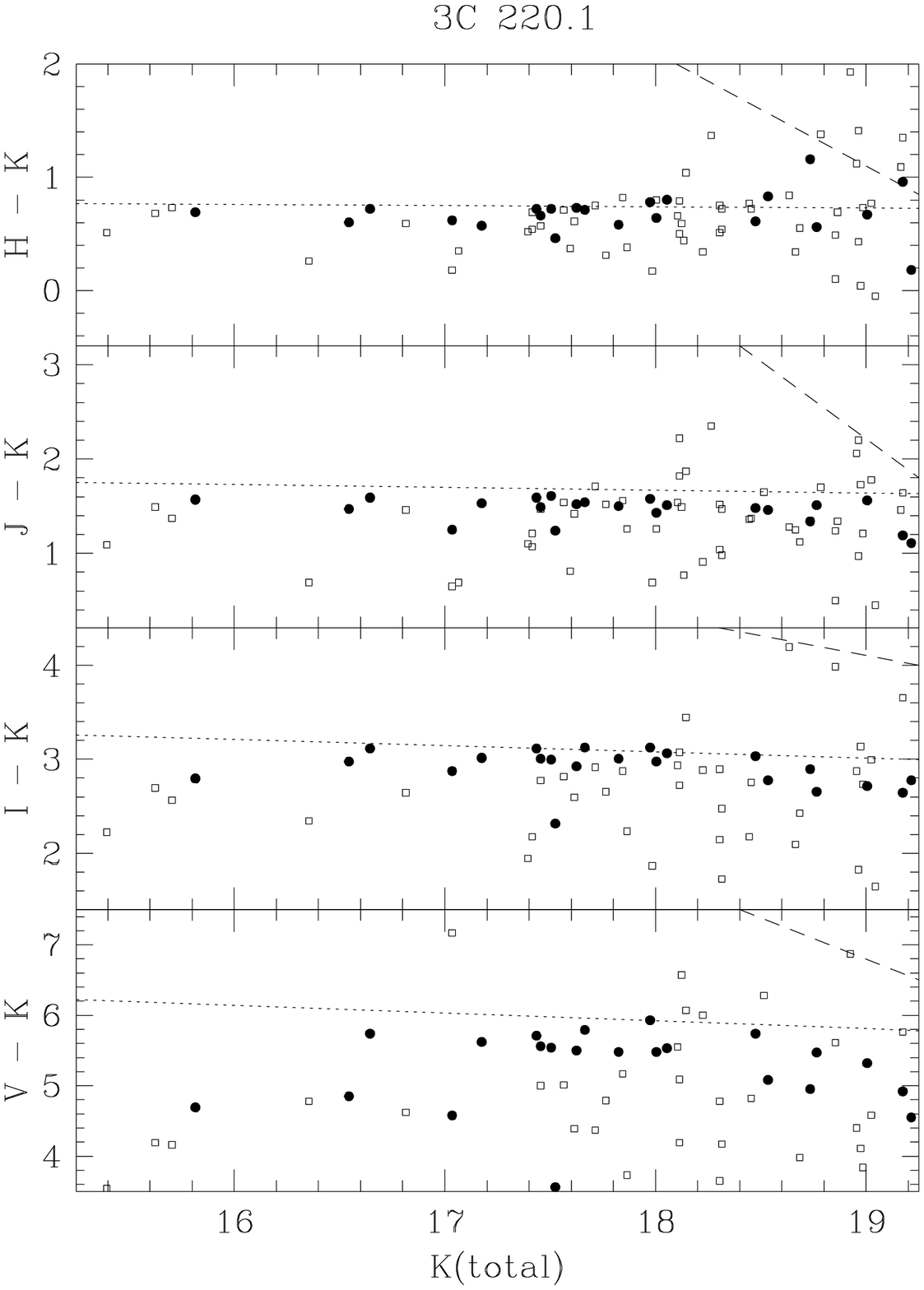}
\caption{}
\end{figure}
\clearpage
\begin{figure}
\figurenum{2n}
\epsscale{0.9}
\plotone{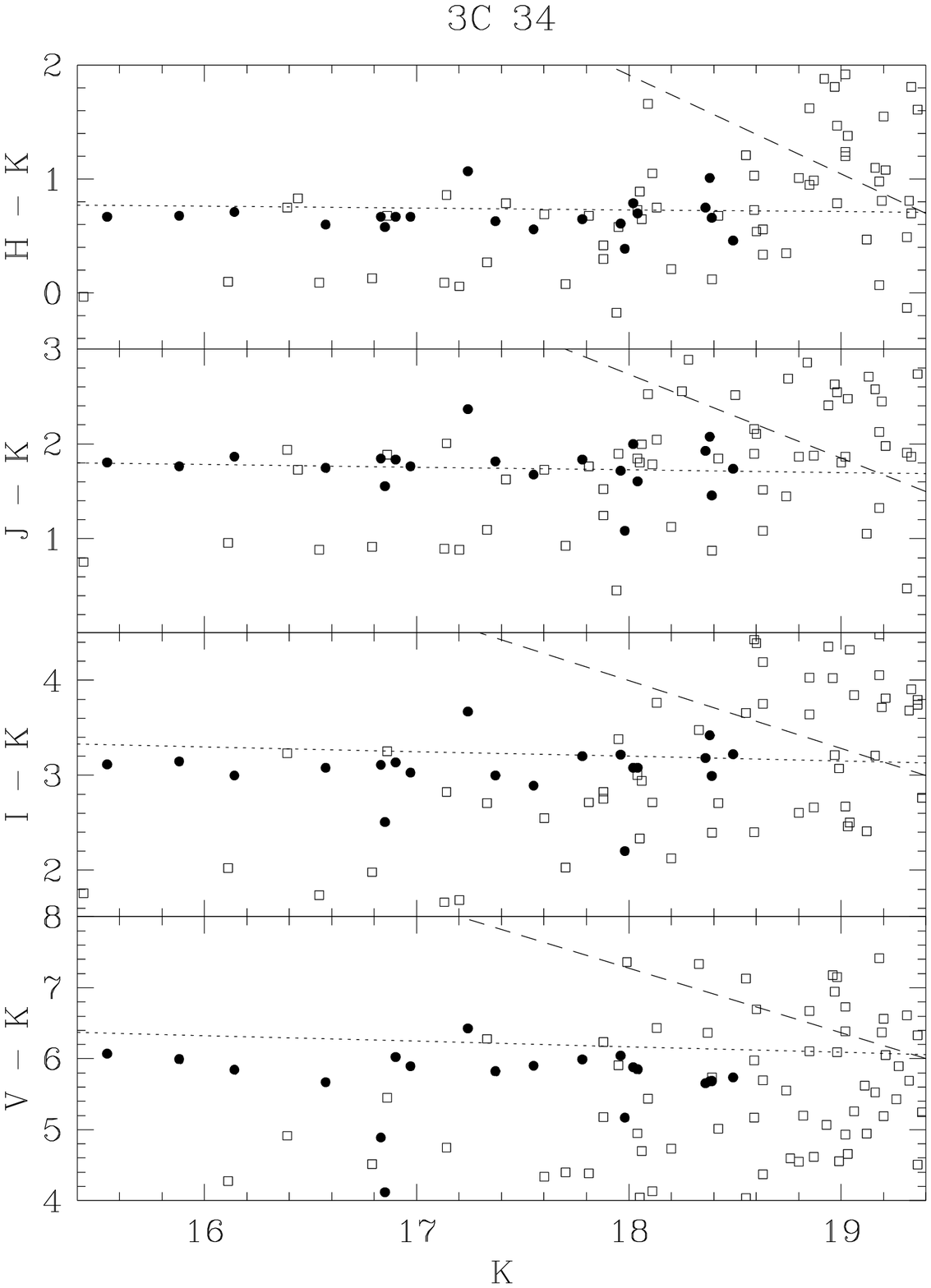}
\caption{}
\end{figure}
\clearpage
\begin{figure}
\figurenum{2o}
\epsscale{0.9}
\plotone{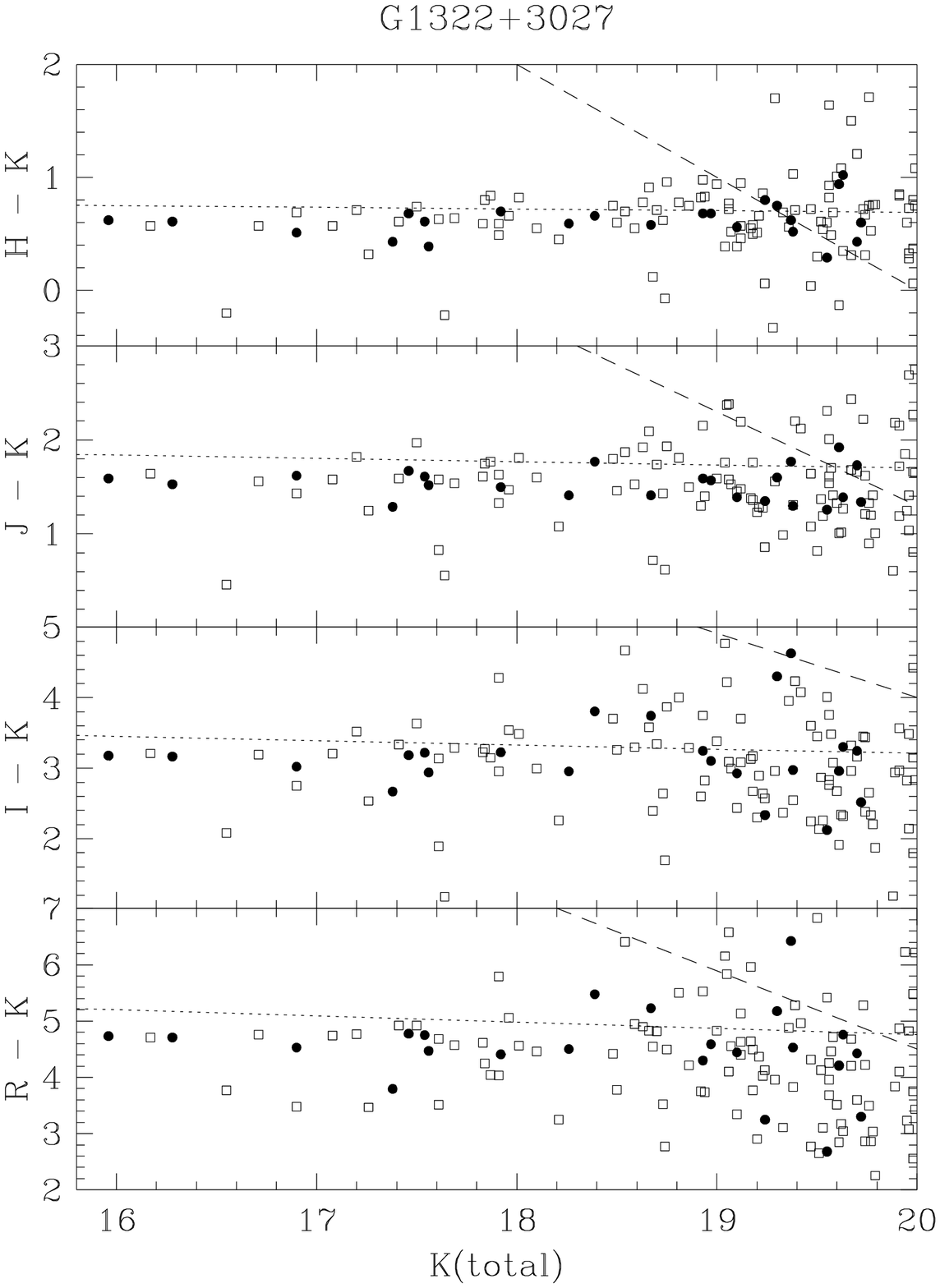}
\caption{}
\end{figure}
\clearpage
\begin{figure}
\figurenum{2p}
\epsscale{0.9}
\plotone{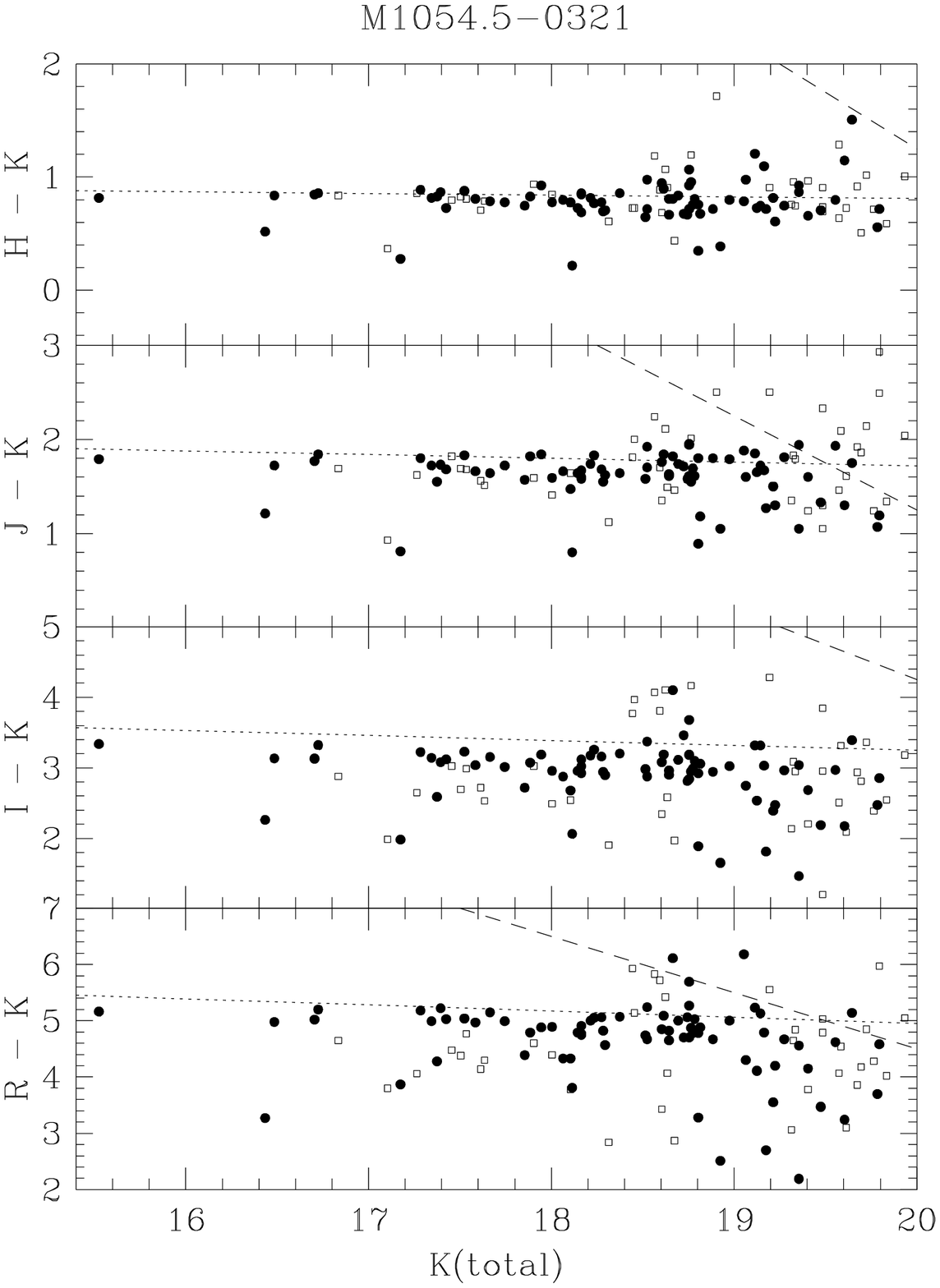}
\caption{}
\end{figure}
\clearpage
\begin{figure}
\figurenum{2q}
\epsscale{0.9}
\plotone{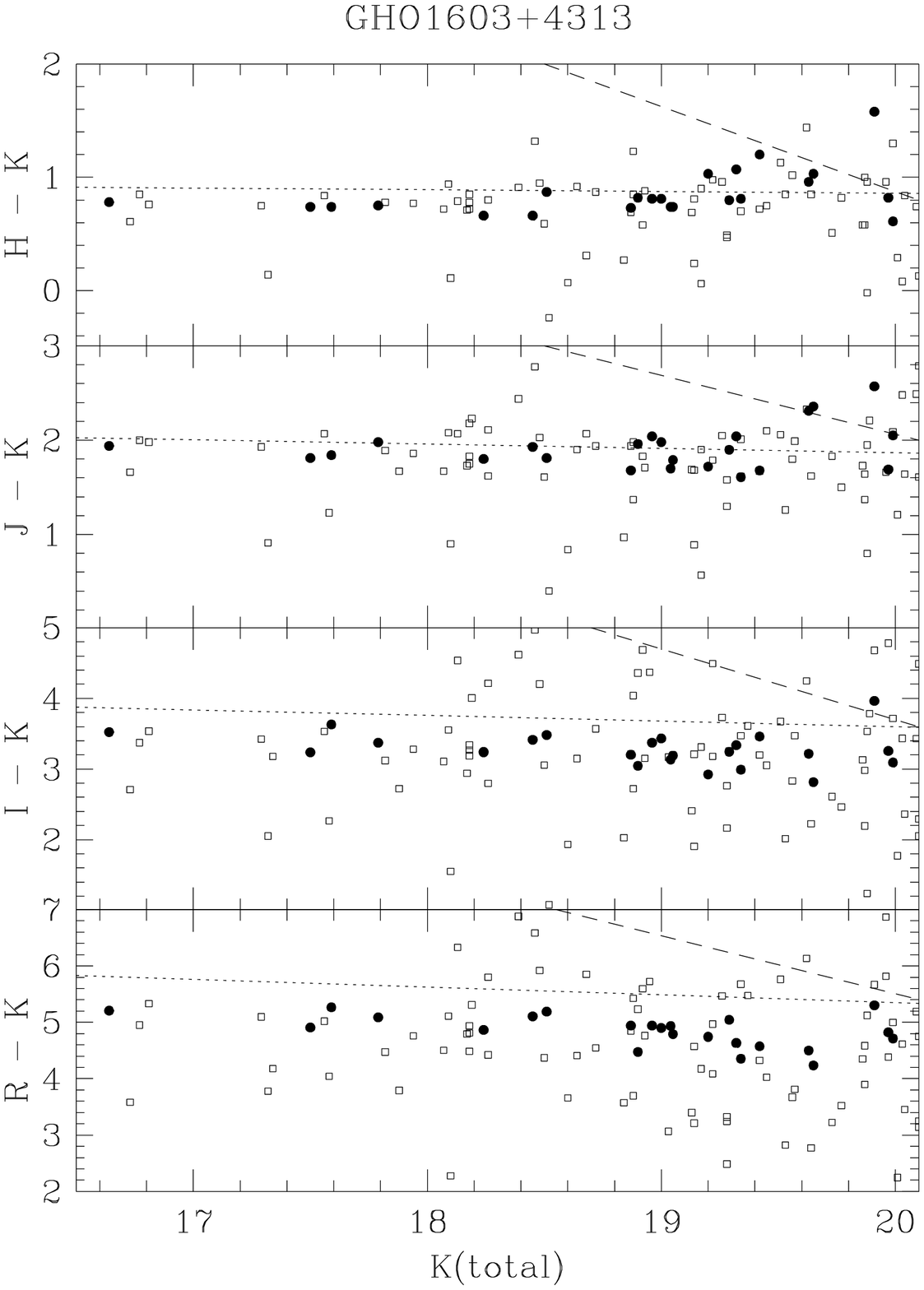}
\caption{}
\end{figure}
\clearpage

\begin{figure}
\figurenum{3}
\epsscale{0.8}
\plotone{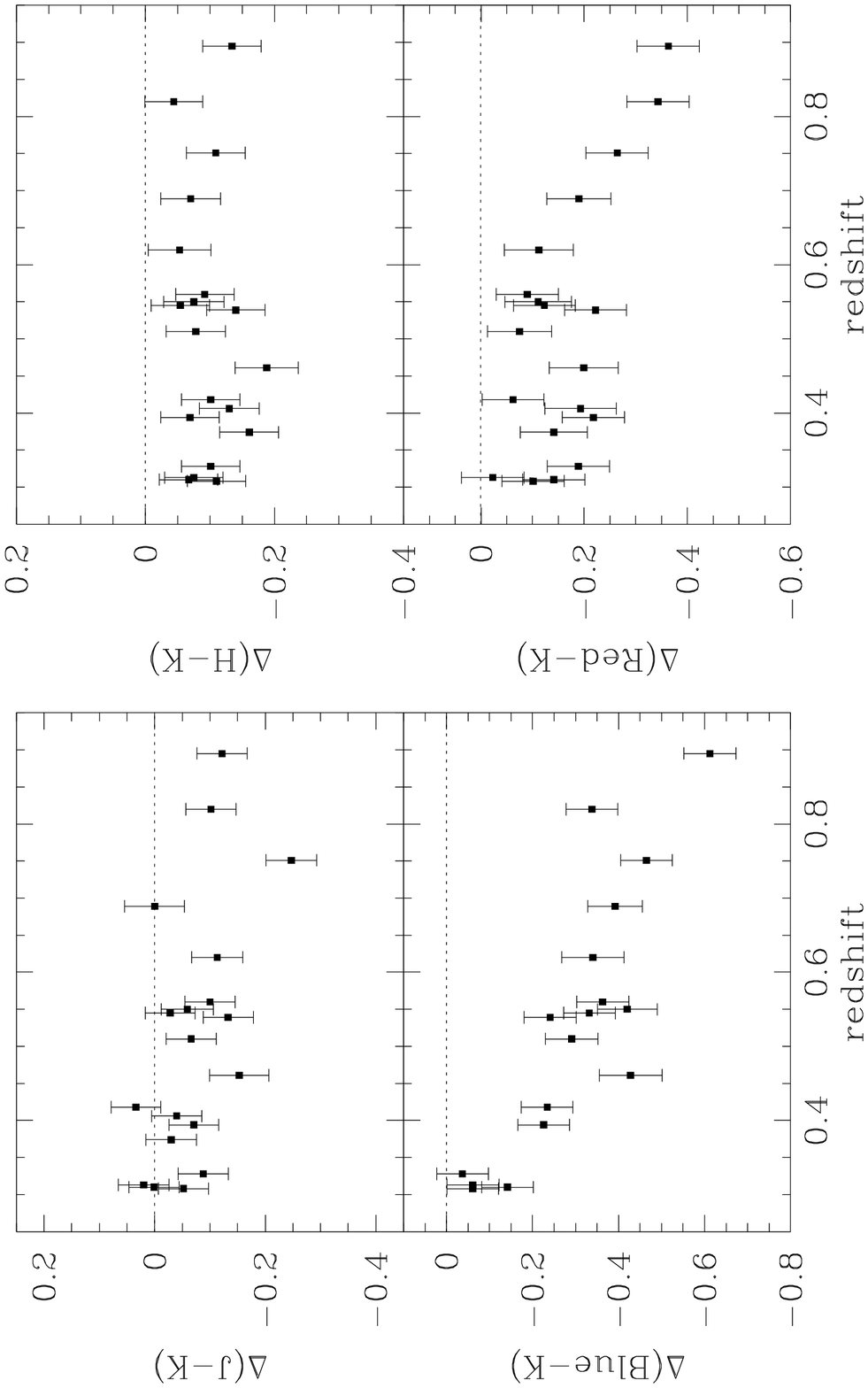}
\caption{Color evolution in early--type galaxies in our distant
cluster sample.  The data points show average differences in the
observed galaxy colors for each cluster relative to the same
rest--frame colors of Coma E+S0 galaxies.  A color difference of
zero (indicated by the horizontal dotted lines) therefore represents
no evolution relative to Coma.  The error bars indicate the
uncertainties in the averages, including our estimate of the
systematic errors.}
\end{figure}

\begin{figure}
\figurenum{4}
\plotone{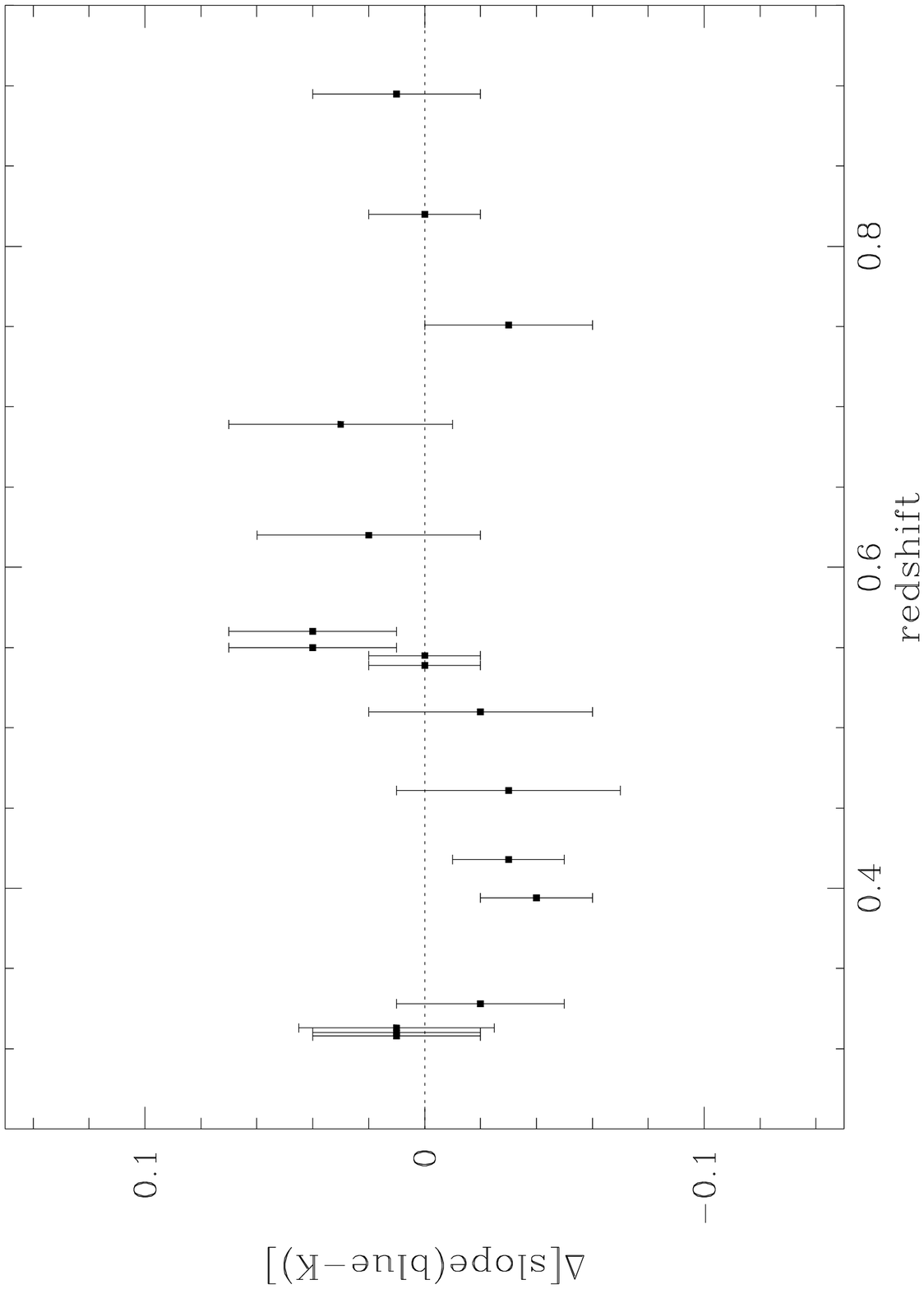}
\caption{The difference between the fitted slopes of the $blue-K$
vs. $K$ color--magnitude relation in each cluster and the corresponding
slopes of the transformed Coma colors, as a function of redshift.  The
dotted horizontal line represents no evolution.  The error bars are
$\pm1 \sigma$ uncertainties. }
\end{figure}

\begin{figure}
\figurenum{5a}
\plotone{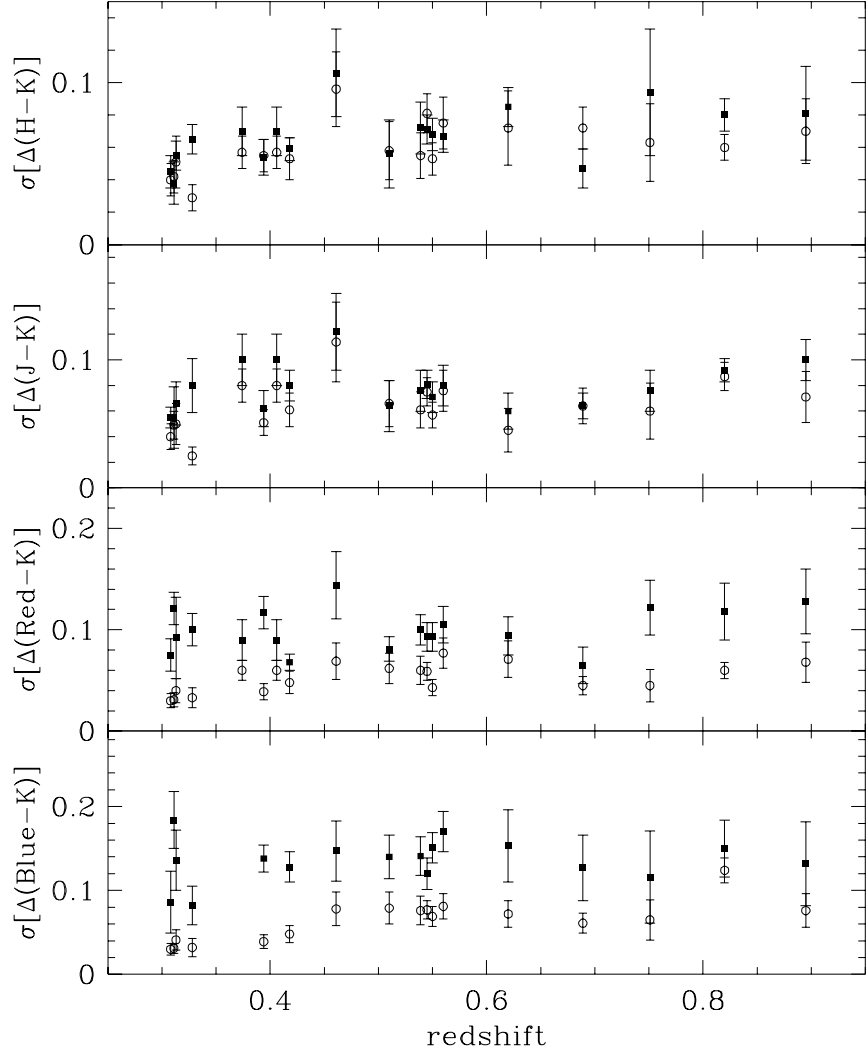}
\caption{The observed (solid squares) and measurement (open
circles) scatter in the colors of early--type galaxies in our cluster
samples.  The error bars show $\pm1 \sigma$ uncertainties in the scatter
values.}
\end{figure}

\begin{figure}
\figurenum{5b}
\plotone{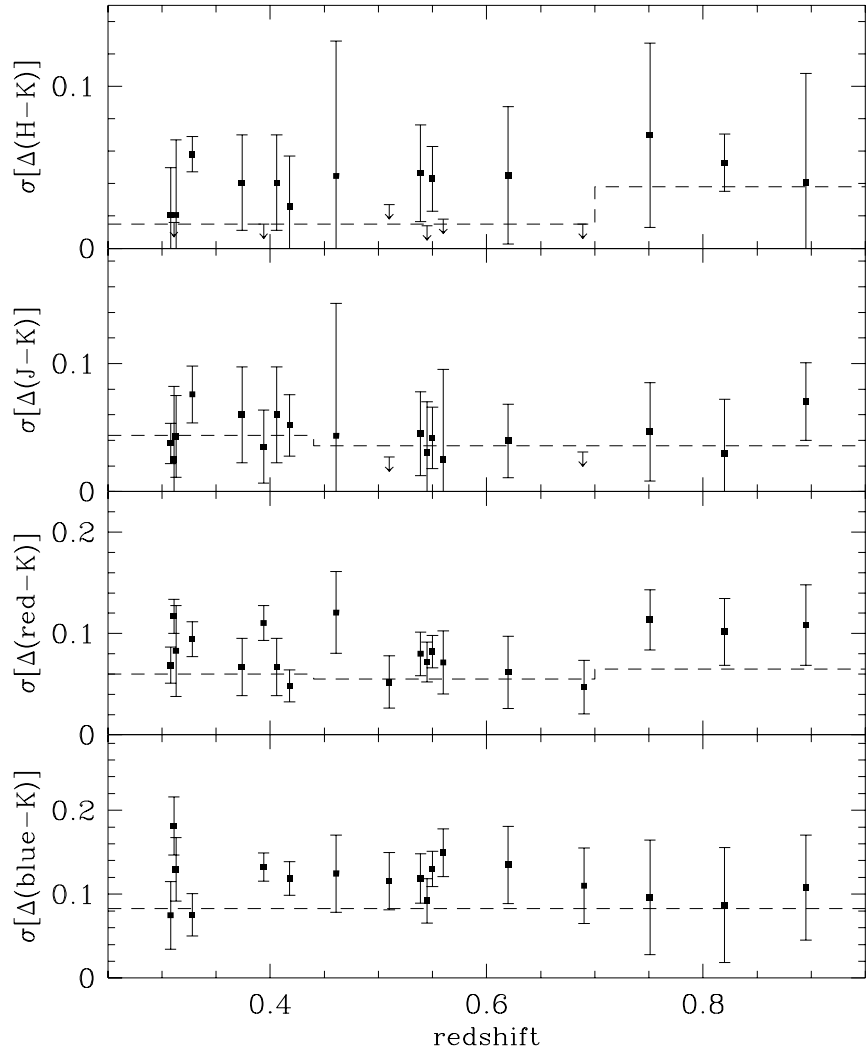}
\caption{Intrinsic scatter in the galaxy colors for each cluster.
The dashed lines represent the intrinsic scatter in similar rest frame
colors for a sample of Coma E+S0 galaxies.  The error bars are
$\pm1 \sigma$ uncertainties in the scatter values.}
\end{figure}

\begin{figure}
\figurenum{6}
\plotone{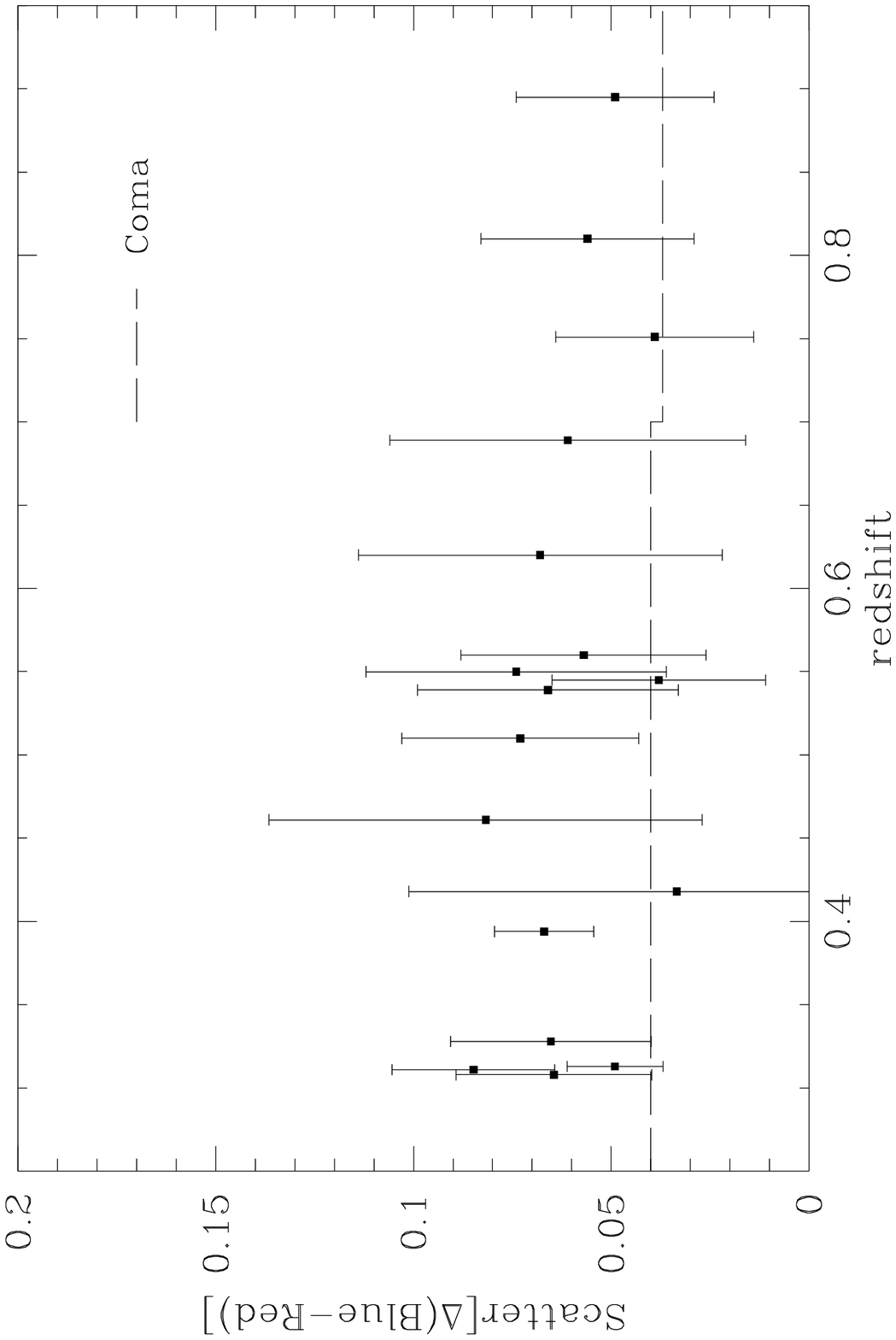}
\caption{The intrinsic scatter in the purely optical $blue-red$ colors 
of the early--type galaxies in each cluster with error bars representing 
$\pm1 \sigma$ uncertainties.}
\end{figure}

\begin{figure}
\figurenum{7}
\epsscale{0.8}
\plotone{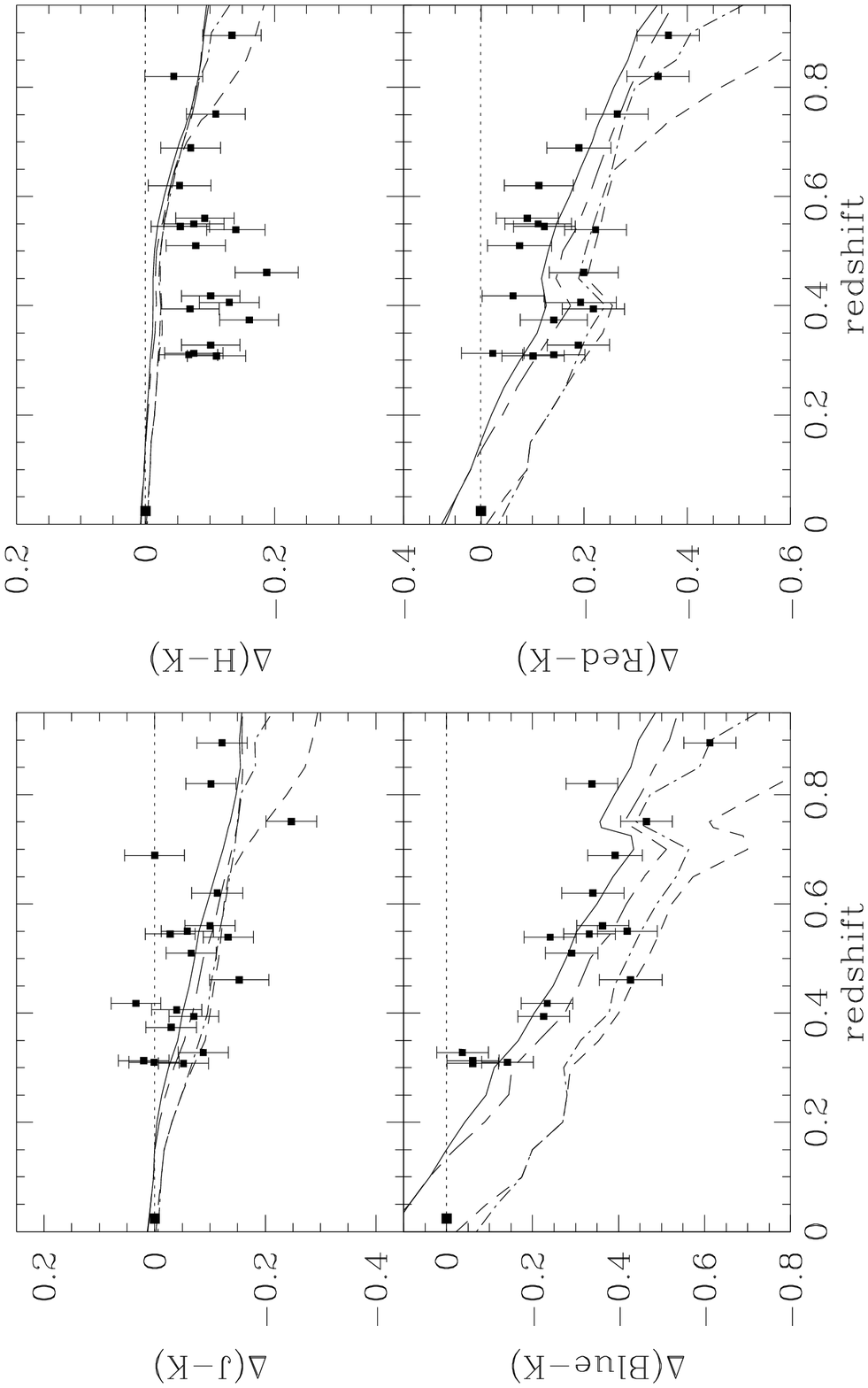}
\caption{Color evolution in distant cluster E+S0s, as presented in 
Figure 3 but with the redshift axis extended to $z = 0$ to include Coma, 
which by definition falls at zero color difference and is represented by 
the points with no error bars.  In each panel, four Bruzual \& Charlot (1996) 
evolutionary synthesis models are plotted.  Each line follows the evolution of 
a 1 Gyr burst of star formation with solar metallicity, plotted with the 
following variations: $z_f = 5$, $h=0.65$ \& $q_0 = 0.05$ (solid line); $z_f=2$, 
$h=0.65$ \& $q_0 = 0.05$ (short dash line); $z_f=5$, $h=0.5$ \& $q_0 = 0.5$ 
(short dash--long dash line); $z_f = 5$, $h=0.8$ \& $q_0 = 0.1$ (dot--dash 
line).  The inflection in the $blue - K$ model colors at $z \approx 0.7$ 
is due to the change of filters used for the observations from the $V$ to $R$
band at this redshift.}
\end{figure}

\newpage

\begin{deluxetable}{lccccccccc}
\small
\tablecaption{Cluster Sample}
\tablewidth{7.1in}
\tablehead{
\colhead{Name} & \colhead{R.A.} & \colhead{Dec.}
& \colhead{$z$} & \colhead{Observed} & \colhead{$K_{lim}$} & \colhead{N$_{samp}$} 
& \colhead{N$_{samp}$} & \colhead{N$_{pred}$} & \colhead{N$_{pred}$}
\\
\colhead{} & \colhead{J2000} & \colhead{J2000}
& \colhead{} & \colhead{bands} & \colhead{mag} & \colhead{total} 
& \colhead{E/S0} & \colhead{field} & \colhead{E/S0}
}
\startdata
AC 118	& 00:14:19.30 & $-$30:23:18 & 0.308 & $gRJHK_s$ & 17.6 & 76 & 38 & 12 & 3 \nl
AC 103	& 20:57:07.46 & $-$64:38:53 & 0.311 & $gRJHK_s$ & 17.6 & 72 & 32 & 12 & 3 \nl
MS 2137.3-234 & 21:40:14.50 & $-$23:39:41&0.313&$gRJHK_s$&17.6&34\tablenotemark{a}&21&12&3\nl
Cl 2244-02 & 22:47:12.90 & $-$02:05:40&0.330&$gRJHK_s$&17.7&43\tablenotemark{a}&24&15&3\nl
Abell 370 & 02:39:53.81 & $-$01:34:24 & 0.374 & $RJHK$ & 17.9 & 79 & 52 & 18 & 4 \nl
Cl 0024+16 & 00:26:35.42 & $+$17:09:51 & 0.391 & $gRJHK_s$ & 18.1 & 90 & 39 & 21 & 4 \nl 
Abell 851 & 09:43:02.60 & $+$46:58:37 & 0.405 & $RJHK$ & 18.2 & 71 & 33 & 23 & 4 \nl
GHO 0303+1706 & 03:06:15.91 & $+$17:19:17 & 0.418 & $gRJHK_s$ & 18.3 & 80 & 38 & 25 & 5 \nl
3C~295	& 14:11:19.47 & $+$52:12:21 & 0.461 & $VIJHK_s$ & 18.5 & 63 & 25 & 28 & 5 \nl
F1557.19TC & 04:12:51.65 & $-$65:50:17 & 0.510 & $VIJHK_s$ & 18.7 & 62 & 29 & 32 & 6 \nl
GHO 1601+4253 & 16:03:10.55 & $+$42:45:35 & 0.539 & $VIJHK_s$ & 18.9 & 79 & 42 & 38 & 8 \nl
MS 0451.6-0306 & 04:54:10.81 & $-$03:00:57 & 0.539 & $VIJHK_s$ & 18.9 & 101 & 51 & 38& 8 \nl
Cl 0016+16 & 00:18:33.64 & $+$16:25:46 & 0.545 & $VIJHK_s$ & 18.9 & 121 & 65 & 38 & 8 \nl
J1888.16CL & 00:56:54.59 & $-$27:40:31 & 0.560 & $VIJHK_s$ & 19.0 & 70 & 38 & 40 & 10 \nl
3C 220.1&09:32:39.61&$+$79:06:32&0.620&$VIJHK_s$\tablenotemark{b}&19.2&59&22&49&12 \nl
3C 34 &01:10:18.53&$+$31:47:20&0.689&$VIJHK_s$&18.5\tablenotemark{c}&49&19&28&7 \nl 
GHO 1322+3027 & 13:24:49.29 & $+$30:11:28 & 0.751 & $RiJHK_s$ & 19.6 & 92 & 23 & 64 & 14 \nl 
MS 1054.5-032 & 10:56:59.53 & $-$03:37:28 & 0.828 & $RiJHK_s$ & 19.8 & 112 & 71 & 73 &15\nl
GHO 1603+4313 & 16:04:18.89 & $+$43:04:36 & 0.895 & $RIJHK_s$ & 20.0 & 95 & 23 & 82 & 16\nl
\enddata
\tablenotetext{a}{WFPC2 image not centered on cluster}
\tablenotetext{b}{V/I photometry from WFPC2 F555W/F814W images}
\tablenotetext{c}{Limiting magnitude only $K^\ast + 1$ due to
relatively shallow WFPC2 image}
\end{deluxetable}

\end{document}